\documentclass[%
reprint,
superscriptaddress,
groupedaddress,
showpacs,preprintnumbers,
aps,
prb,
]{revtex4-2}
\usepackage[utf8]{inputenc}
\usepackage{graphicx}
\usepackage{xr}
\usepackage{mwe}
\usepackage[dvipsnames]{xcolor}
\usepackage{hyperref}
\usepackage{ulem}
\usepackage{amssymb}
\usepackage{wasysym}
\usepackage{rotating}
\begin{document}
   \title{Step meandering: The balance between the potential well and the Ehrlich -- Schwoebel barrier}%
  %\pacs{?,?,?,?,?}
  \author{Marta A. Chabowska$^1$}%
  \email[]{galicka@ifpan.edu.pl}%
  \author{Hristina Popova$^2$}%
  %\email[]{karleva@ipc.bas.bg}
  \author{Magdalena A. Za{\l}uska-Kotur$^1$}%
  %\email[]{zalum@ifpan.edu.pl}
  \affiliation{$^1$Institute of Physics, Polish Academy of Sciences, al. Lotnik\'{o}w 32/46, Warsaw, Poland, \\
  $^2$Institute of Physical Chemistry, Bulgarian Academy of Sciences, Acad. G. Bonchev str., 1113 Sofia, Bulgaria}%

\date{\today}%

\begin{abstract}
This study presents a comprehensive and innovative exploration of how the surface potential energy landscape influences meander formation. Using the Vicinal Cellular Automaton model, which distinguishes surface diffusion from adatom incorporation into the crystal, the research delves into various factors affecting surface pattern dynamics.
By isolating the diffusion process within a defined energy potential, the study provides a detailed analysis of how changes in the potential energy well and the barrier at the top of the step contribute to meander formation. Remarkably, the results reveal that the mere presence of a potential well at the step’s bottom is sufficient to induce meandering.
The role of the Ehrlich-Schwoebel barrier on formed meanders is further investigated, and a mechanism for meander formation is proposed to clarify this process.
The derived relation successfully reflects the wavelength of the meandered patterns observed in the simulations, emphasizing its reliability. Overall, the results illustrate the crucial influence of the surface energy potential's shape in driving surface pattern formation.
\end{abstract}

\maketitle
\section{Introduction}

Surface morphology and its evolution during crystal growth are crucial in the fabrication of electronic and optoelectronic devices. Achieving control over surface growth to obtain a desired surface pattern is a key goal. However, this control is challenging due to surface instability, as even small energy barriers can have significant effects. One of the primary instabilities is step meandering, a morphological instability driven by surface  diffusion. Step meandering has been observed in experiments \cite{Damilano-JAP, Chou-ASS,Wu-JCG,Turski-ASS,Gocalinska-ASS,Pandey-Vac} and explored theoretically in numerous studies \cite{Bales-PRB,OPierre-PRB,OPierre-PRL,MZK-JCG12,Beausoleil-PRE,Blel-PRB}. In general, previous studies have shown that meanders develop with a fixed wavelength and unbounded amplitude during the growth process.

Understanding the mechanisms behind step meander formation is crucial from both theoretical and experimental perspectives. One of the first explanations for this instability, established in subsequent studies, is the role of the Ehrlich -- Schwoebel (ES) barrier. The ES barrier is an additional diffusion obstacle that adatoms encounter when attempting to descend steps. It has been demonstrated to play a crucial role in meander formation during growth processes \cite{Bales-PRB,ESB,ESB-exp99}.

The formation of step meandering has been extensively studied using the kinetic Monte Carlo method, which has shown that the ES barrier, along with factors such as temperature, particle flux \cite{MZK-JCG12}, and nucleation processes \cite{Beausoleil-PRE}, strongly influences this phenomenon. Additionally, Ref. \cite{Hamouda-SS} presents an analysis incorporating the ES barrier, illustrating how the meander wavelength is linked to effective step stiffness and adatom mobility — both of which are significantly impacted by kink concentration. The kink ES barrier as a primary mechanism for the step meandering process was introduced in \cite{OPierre-PRL}.

The relationship between kinks and meander instability under the influence of the kink ES barrier was further validated in Ref. \cite{Kallunki-PRB}. Another critical factor affecting meander development is the strength of next-nearest-neighbor interactions, which influence the correlation between the meander wavelength and the deposition rate \cite{Blel-PRB,Hamouda-SS}. Studies employing highly nonlinear evolution equations \cite{Pierre-PRL,Gillet-EPJ,Paulin-PRL} have also revealed that elastic step interactions contribute to the lateral coarsening of meanders. Numerous mechanisms contributing to step meandering instability have been identified and extensively studied, many of which can be understood as different manifestations of a common underlying source. In our work, we undertake a systematic investigation of this phenomenon, focusing on the perspective of the surface potential experienced by adatoms. These adatoms diffuse across the surface and eventually attach to the crystal, becoming part of its structure and thereby shaping the resulting surface morphology.

The findings regarding the surface potential, obtained by Toru Akiyama and co-authors through \textit{ab-initio} calculations \cite{Ohka-CGD,Akiyama-JCG20,Akiyama-JCG21,Akiyama-JJAP}, are highly significant for our analysis. Their research examined the stability of vicinal surfaces of GaN and AlN and the influence of step edges on particle adsorption behavior. These studies show that particles diffusing on a vicinal surface encounter a varying surface potential dependent on their position. Notably, they calculated the depth of the potential well at the bottom of the step and observed that the ES barrier appears only at certain types of steps.

In this paper, we explore the formation of meanders and investigate their emergence in the absence of the ES barrier. Our approach is based on considering the surface from the perspective of  the potential at the surface and afterward examining the effect of this potential on the surface morphology. We further investigate the influence of the presence of the Ehrlich -- Schwoebel barrier on meanders formation. The model used here is based on a new approach to the previously introduced (2~+~1)D vicinal Cellular Automaton (VicCA) model \cite{MZK-crystals,Chabowska-ACS}. Our findings demonstrate that the ratio of adatom attachment at the kink position to that at the step is a critical factor in the formation of meanders at the surface, particularly in the presence of a potential well at the bottom of the step.

%%%%%%%%%%%%%%%%%%%%%%%%%%%%%%%%%%%%%%%%%%%%%%%%%%%%%%%%%%%%%%%%%%%%%%%%
%                                Model
%%%%%%%%%%%%%%%%%%%%%%%%%%%%%%%%%%%%%%%%%%%%%%%%%%%%%%%%%%%%%%%%%%%%%%%%
\section{The model}
Our model, a new approach to (2~+~1)D vicinal Cellular Automaton model, simulates the evolution of a crystal surface and the diffusion of adatoms on it. Previous studies have explored this model in (1~+~1)D \cite{Krasteva-AIP,Krzyzew-JCG,Toktarbaiuly-prb,Krzyzew-CGD,Popova-CGD,Popova-CGD23} and (2~+~1)D \cite{MZK-crystals,Chabowska-ACS,Chabowska-Vac} context. The model combines a Cellular Automaton (CA) module for surface growth and a Monte Carlo (MC) module for adatoms diffusion. The CA module updates the surface in parallel based on pre-defined rules, while the MC module simulates adatoms diffusion sequentially.

The model comprises two main components: the crystal surface and a layer of adatoms. Adatom diffusion is governed by a standard  MC procedure. Adatoms are randomly selected and move to adjacent empty sites with a probability determined by the energy potential of the site. A single diffusion step is considered complete when each adatom has been visited, moved once on average. This approach allows for large-scale simulations. Between two growth modules, all adatoms try $n_{DS}$ diffusional jumps. As $n_{DS}$ grows, the growth mode changes from diffusion-limited (DL) to kinetics-limited (KL), and at the same time the transparency of the step increases.

\begin{figure*}[hbt]
 \centering
\includegraphics[width=0.3\textwidth]{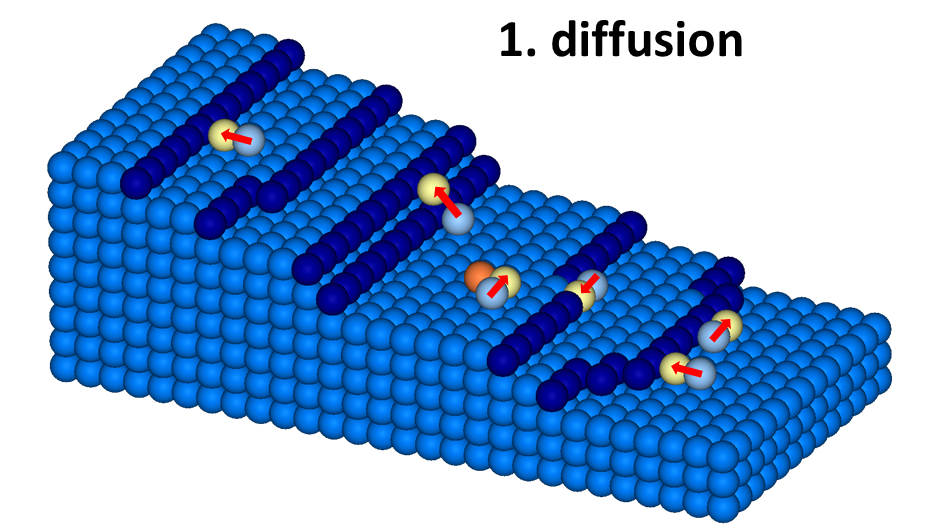}
\includegraphics[width=0.3\textwidth]{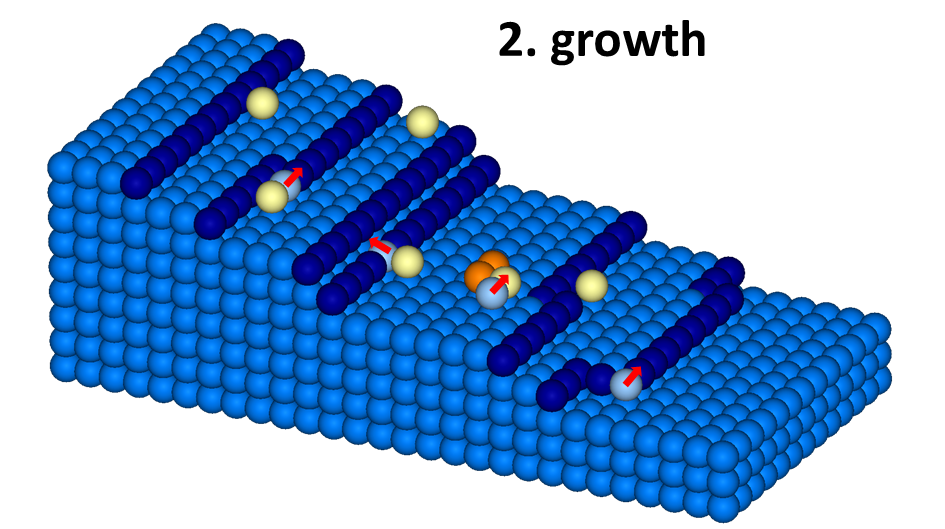}
\includegraphics[width=0.3\textwidth]{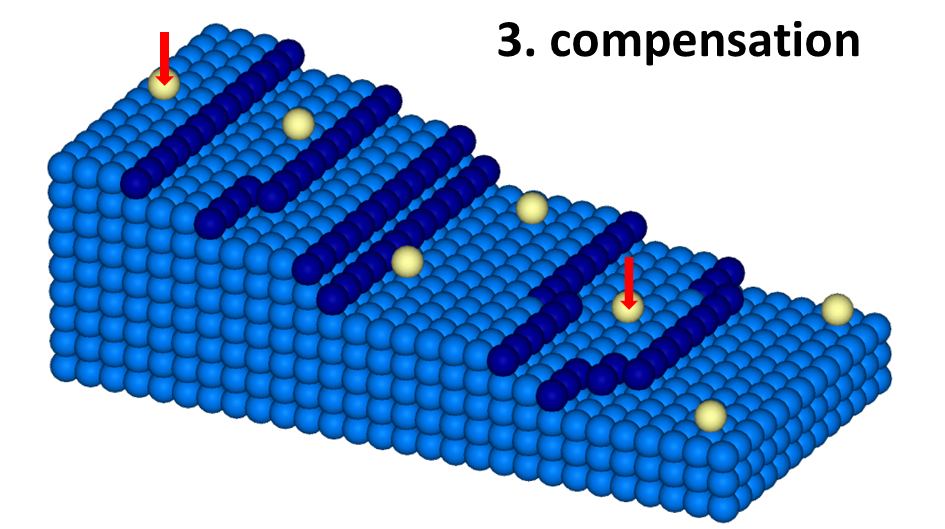}
\caption{A single time step of the simulation procedure consists of the following stages: (1) the diffusion process, (2) the growth update, and (3) the adatom concentration compensation. For the system shown, the number of adatoms must be adjusted to 7 in the last stage of the time step to keep the initial concentration of adatoms constant.}
\label{fig:procedure}
\end{figure*}

The crystal part of  the  model  is build  on  a square  lattice. The surface typically consists of steps descending from left to right,   with the initial distance between them being  $l_0$. Periodic boundary conditions are applied along the steps, while helical periodic boundary conditions are applied across the steps to maintain the height differences. The incorporation of adatoms into the crystal is governed by CA rules. Three distinct scenarios have been identified in which an adatom becomes part of the crystal structure. The most common locations where an adatom attaches to a crystal and is immediately incorporated are the kinks that are present at the corners of the step. This attachment is determined by the probability, designated as $P_k$. In the second scenario, an adatom situated in close contact with both a straight step and another adatom becomes a potential crystal site. This incorporation occurs with a probability of $P_s$. These rules determine the stiffness of the step. We assumed that particles are easily built in the crystal at kinks, and more difficult at the straight part of the step. The step stiffness can be regulated by increasing or decreasing the probability of the second event. The stiffness of the step is increased when it is more difficult to incorporate adatoms into a straight step. A third situation in which the adatom becomes part of the crystal layer is also possible; in this case, the adatom becomes the nucleus of the new layer regardless of the position of the step. This occurs when at least four adatoms stick together on the terrace. More details about the model can be found in Ref.~\cite{MZK-crystals}. The simulation time step consists of three key processes: the diffusion of adatoms across the surface, updating the surface growth within the  CA module, and the compensation of adatoms to their initial concentration, $c_0$ (see Figure~\ref{fig:procedure}). The compensation ensures that the growth process continues by introducing new atoms into the system. By doing so, a constant incoming flux is effectively simulated, where the flux value is proportional to  $c_0$.

\begin{figure*}[hbt]
 \centering
a)\includegraphics[width=0.3\textwidth]{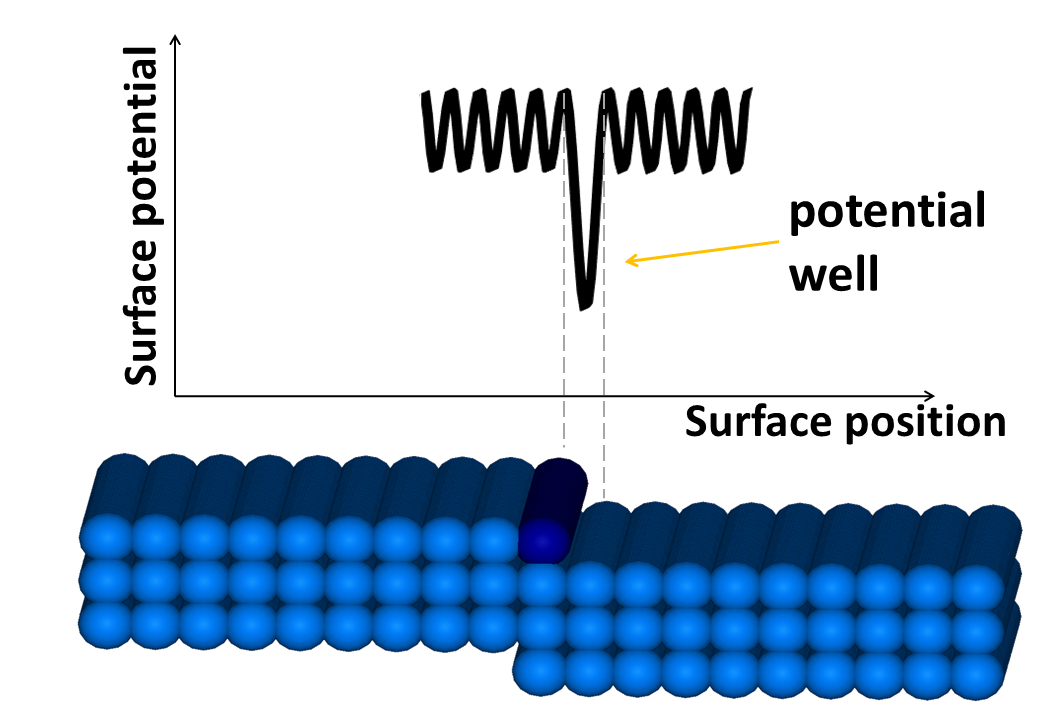}
b)\includegraphics[width=0.3\textwidth]{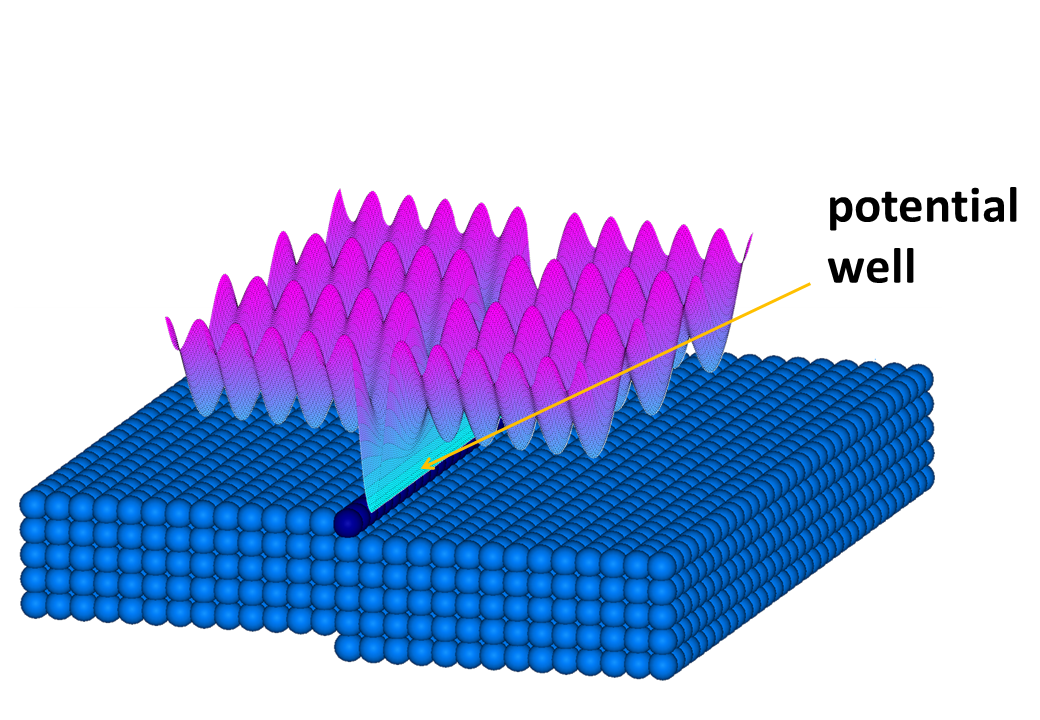}
c)\includegraphics[width=0.3\textwidth]{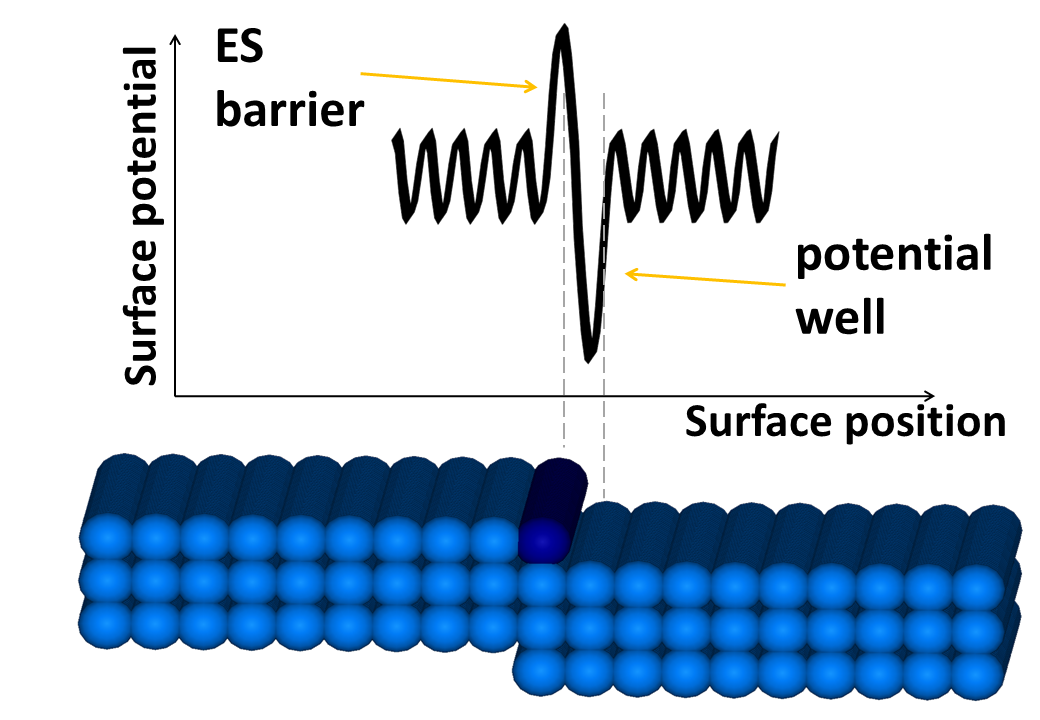}
\caption{Visualization of the potential well at the bottom of the step. View from a) the side, b) the top at an angle, c) the side with the additional Ehrlich -- Schwoebel barrier}
\label{fig:potential}
\end{figure*}

The most significant aspect of the model refers to the diffusion of adatoms which serves as the driving force behind the formation of surface patterns. The diffusion process is critically dependent on the energy potential landscape experienced by the atoms.  Based on the knowledge derived from the \textit{ab-initio} results \cite{Akiyama-JCG21} we postulate the presence of a potential well (Figure~\ref{fig:potential}a,b) at the bottom of the step, along it, similarly as shown in Ref.\cite{Blel-VAC}. It was assumed that the depth of the potential well remains constant throughout the duration of the simulation, equal to $ E_V$, the energy associated with the potential well. However, the shape of the surface potential is influenced by a number of different factors, including the presence of dangling bonds, surface reconstruction, and the existence of additional barriers. From the perspective of meander analysis, the previously mentioned Ehrlich -- Schwoebel barrier at the top of the step is of particular significance. In the event that such a barrier is present, the shape of the potential in our model assumes the form depicted in Figure~\ref{fig:potential}c. The presence of this barrier makes it more challenging for particles to execute jumps across the step. The height of this barrier is designated as $ E_{ES}$. In the model, it is assumed that each particle diffuses independently. All jumps along terraces, with the exception of those in the immediate vicinity of the step, are performed with the same probability. The  probability  of each diffusional jump depends on the surface diffusion barrier $E_D$. Probability $P_0=\exp(-\beta E_D)$  depends on temperature factor $\beta=1/k_B T$ with T denoting temperature and $k_B$ representing the Boltzmann constant. To speed up the process we divide all jump probabilities by $P_0$, thus increasing the probability of each simple jump along the surface to 1 after the equal choice of jump direction.
Thus, the probability of a jump  is equal to
\begin{equation}
\frac{P_V}{P_0} = \left\{\begin{array}{rcl}
&e^{-\beta E_V} & \ \text{out}\\
&1 & \ \text{in}
\end{array} \right.
\end{equation}
for jumps  out and in  the  well accordingly,
whereas in the presence of an ES barrier, this probability of a jump across the steps is set to
\begin{equation}
\frac{P_{ES}}{P_0} = \left\{\begin{array}{rcl}
& e^{-(\beta E_{ES}+\beta E_V)}& \ \text{out}\\
& e^{-\beta E_{ES}} & \ \text{in}
\end{array} \right.
\end{equation}
The potential well plays a role in localizing particles, resulting in an increase in particle density at the bottom of the step. We should note that in our computer simulations all potential energies are measured in units of thermal energy $k_B T$, or the potentials of interest are given by $\beta E_V$ and $\beta E_{ES}$.

%%%%%%%%%%%%%%%%%%%%%%%%%%%%%%%%%%%%%%%%%%%%%%%%%%%%%%%%%%%%%%%%%%%%%%%%
%                                Results
%%%%%%%%%%%%%%%%%%%%%%%%%%%%%%%%%%%%%%%%%%%%%%%%%%%%%%%%%%%%%%%%%%%%%%%%
\section{Results and discussion}
The primary objective of this paper is to present a clear and simple mechanism for the formation of meanders during the initial stages of their development, both in the absence and presence of the Ehrlich -- Schwoebel barrier.  In light of the findings pertaining to the surface potential of diffusing adatoms on crystalline surfaces \cite{Ohka-CGD,Akiyama-JCG21,Akiyama-JJAP,Akiyama-JCG20}, we investigated the diverse forms of surface potential and their impact on the formation of the final surface pattern. Furthermore, our aim is to demonstrate which structural behaviors are attributable to the various components of the surface potential. Our investigations were conducted  for varying depths of the applied surface potential well and height of the Ehrlich -- Schwoebel barrier, if present. The resulting patterns demonstrated stability throughout the process, predominantly assuming a wave-like configuration, namely meanders.

%%%%%%%%%%%%%%%%%%%%%%%%%%%%%%%%%%%%%%%%%%%%%%%%%%%%%%%%%%
%%%%%           Potential well                       %%%%%
%%%%%%%%%%%%%%%%%%%%%%%%%%%%%%%%%%%%%%%%%%%%%%%%%%%%%%%%%%
\subsection{Potential well}
\begin{figure*}[hbt]
 \centering
a)\includegraphics[width=0.22\textwidth]{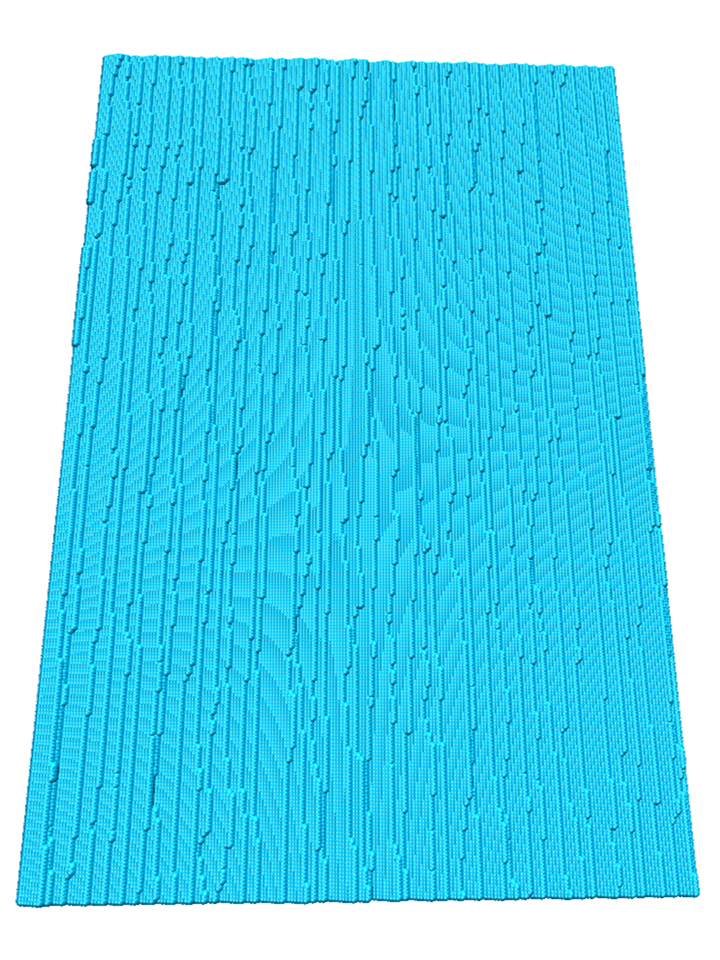}
b)\includegraphics[width=0.22\textwidth]{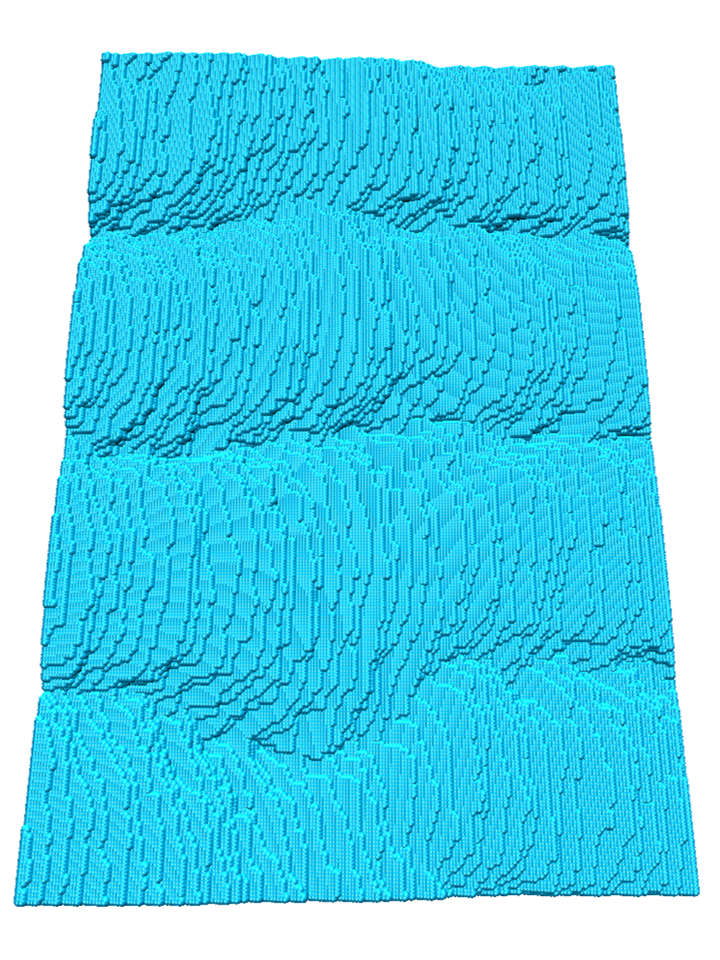}
c)\includegraphics[width=0.22\textwidth]{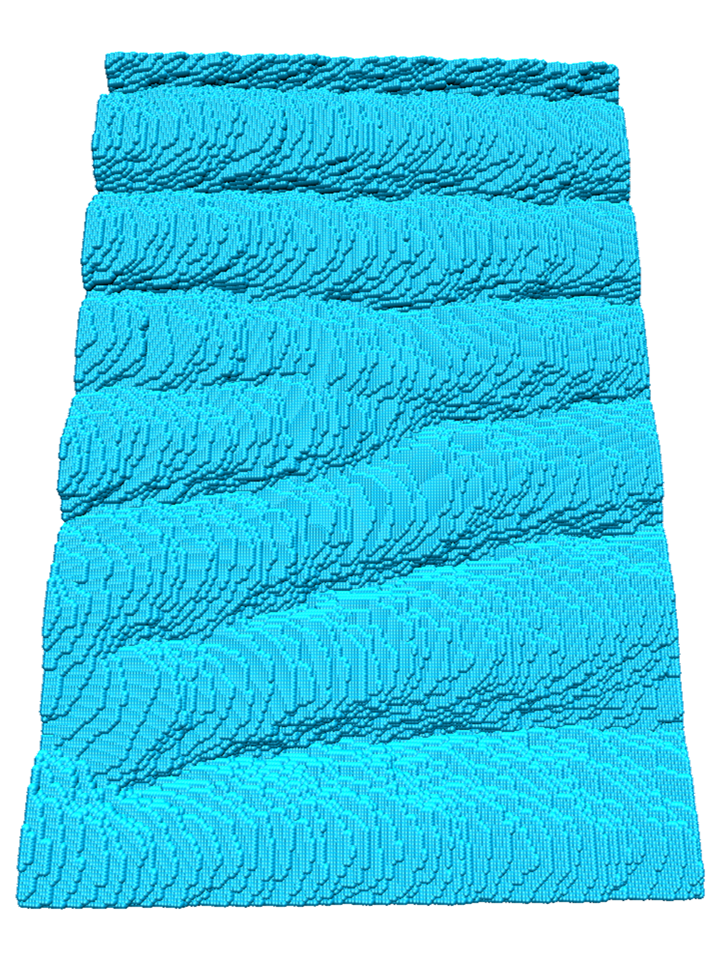}
d)\includegraphics[width=0.22\textwidth]{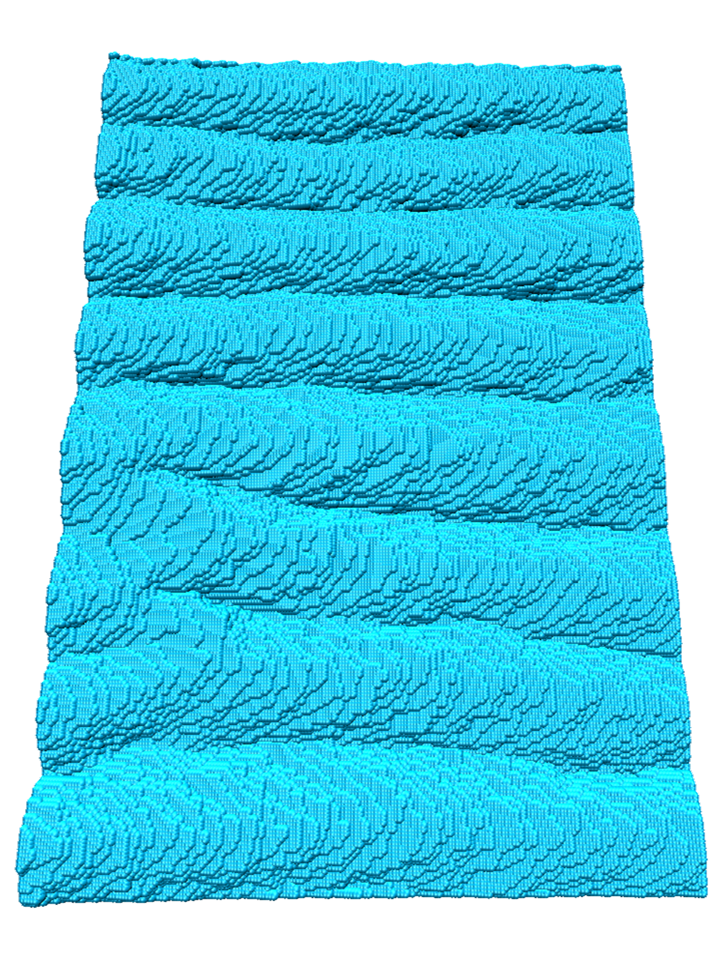}
\caption{Meanders obtained for $c_0 = 0.003$, $l_0 = 5$, $n_{DS} = 5$ and a) $\beta E_V = 0.0$, b) $\beta E_V = 2.0$, c) $\beta E_V = 3.5$, d) $\beta E_V = 6.0$, simulation time $t = 2 \cdot 10^6$. System size $200$ x $300$.}
\label{fig:pot_wel-dep}
\end{figure*}

The study of meanders begins with an analysis of the influence of the depth of the surface potential well on the formation of meanders in the absence of an Ehrlich -- Schwoebel barrier. The depth of the potential well, $\beta E_V$, was considered within a range of $0.0$ to $10.0$. This corresponds to the probability of a particle jumping out of the well, ranging from $1$ (the jump will occur) to $4.5 \cdot 10^{-5}$ (the jump is less likely to occur). Factor $\beta E_V = 8.0$ means that at temperature of about $700~K$ the energy is $0.48$~eV. The results obtained after $2 \cdot 10^6$ VicCA simulation time steps are presented in Figure~\ref{fig:pot_wel-dep} for selected values of  $\beta E_V$: $0.0$, $2.0$, $3.5$ and $6.0$. Initially, the result concerning the scenario without a potential well are presented (Figure~\ref{fig:pot_wel-dep}a).  In this case, the final surface configuration exhibits a regular step-like ordering. The introduction of a potential well at the bottom of the step results in the emergence of meanders with a wavelength $\lambda$ (Figure~\ref{fig:pot_wel-dep}b). An increase in the depth of the potential results in a reduction in the wavelength of the meanders, as illustrated in Figures~\ref{fig:pot_wel-dep}c and \ref{fig:pot_wel-dep}d. From a certain depth of the potential well ($\beta E_V \approx 5.5$), the wavelength $\lambda$ does not decrease with the further increasing of the depth of potential well. This can be explained by the fact that the density of particles at step does not increase any more even if potential well is deeper. Additionally, we verified that the meander wavelength is independent of the system size by testing it on systems four  times larger (See Fig. S2 in Supplementary Material). Furthermore, we observed that once the meanders become well-formed and stable, their wavelength remains constant, even when the simulation time was extended to 10 times its original duration (See Fig. S3 in Supplementary Material).

Certainly, the meandering process is influenced by multiple parameters within the model, each capable of altering the system's behavior. One of these parameters is the initial terrace width $l_0$, which, in experimental contexts, corresponds to the so-called miscut angle — essentially, the angle at which the crystal is cut.

\begin{figure}[hbt]
 \centering
\includegraphics[width=0.5\textwidth]{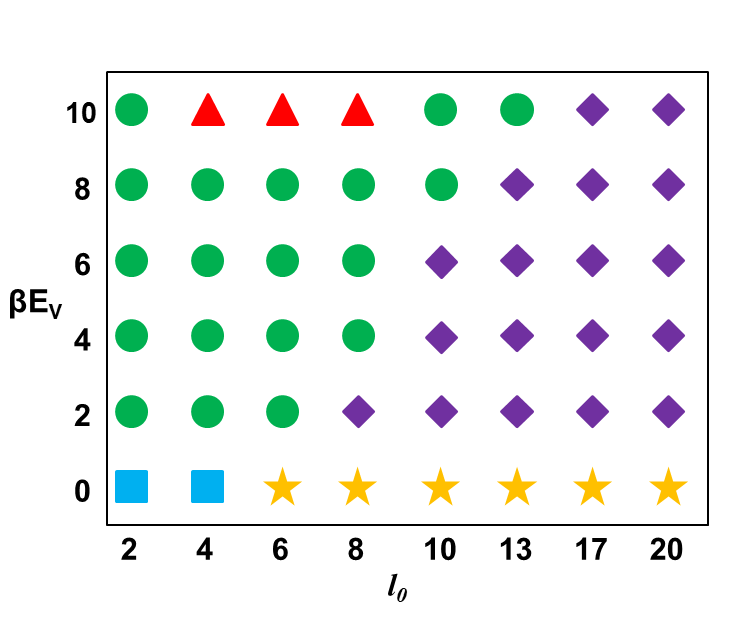}
\caption{Diagram of pattern formation on the surface obtained for  $c_0 = 0.003$, $\beta E_{ES} = 0$, $n_{DS} = 2$ and $t = 1 \cdot 10^6$. Different symbols correspond to different patterns:  {\color{blue}$\blacksquare$} describes regular step ordering, {\color{YellowOrange}$\bigstar$} means regular step ordering with islands, {\color{ForestGreen}\CIRCLE} describe meanders formed only by the movement of kinks, {\color{violet}$\blacklozenge$} describe meanders formed by the movement of kinks and the islands appearing on the terraces and {\color{red}$\blacktriangle$} means rough surface.}
\label{fig:diagram_V-0}
\end{figure}

In Figure~\ref{fig:diagram_V-0}, we illustrate how the formation of surface patterns evolves with changes in both the depth of the potential well and the terrace width $l_0$. This visualization highlights the interplay between these two factors in shaping the resulting surface structure. The diagram illustrates four distinct surface structures arising under different conditions. When the potential well is zero, steps maintain their regular, parallel arrangement. However, from certain $l_0$, islands form on the terraces and eventually attach to the steps, which then straighten as they absorb the islands. When the potential well reaches $2k_BT$ or higher, meanders start to form. These meanders can result either from steps bending on their own or from a combination of step bending and the attachment of islands formed on wider terraces. As the potential increases further, the likelihood of island formation decreases. This is because higher potential values trap more particles at the steps, reducing particle density on the terraces. At very high potential values and small $l_0$, regular meandering patterns disappear. Instead, step bending leads to irregular and disordered structures. It is important to note that in all these structures, the islands remain at a single-layer height, resulting in relatively flat surface formations. The emergence of three-dimensional structures is more frequently observed when  the ES barrier  is  present, as it promotes nucleation on  top  of the islands. In the model, the probability of island creation can be adjusted by modifying the CA rules accordingly. Specifically, it is possible to entirely prevent the formation of islands. In the Supplementary Material (Fig. S4), we present a comparison of the meandering process with and without the possibility of nucleation in several representative cases.

%%%%%%%%%%%%%%%%%%%%%%%%%%%%%%%%%%%%%%%%%%%%%%%%%%%%%%%%%%
%%%%%           Ehrlich-Schwoebel barrier            %%%%%
%%%%%%%%%%%%%%%%%%%%%%%%%%%%%%%%%%%%%%%%%%%%%%%%%%%%%%%%%%
\subsection{The Ehrlich -- Schwoebel barrier}
\begin{figure*}[hbt]
 \centering
a)\includegraphics[width=0.22\textwidth]{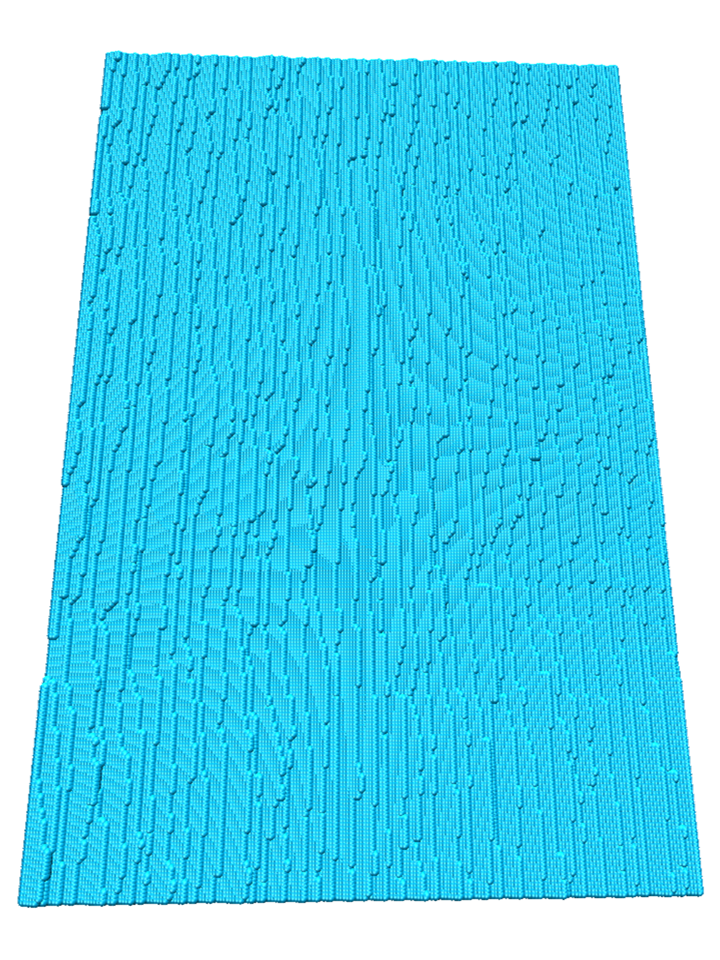}
b)\includegraphics[width=0.22\textwidth]{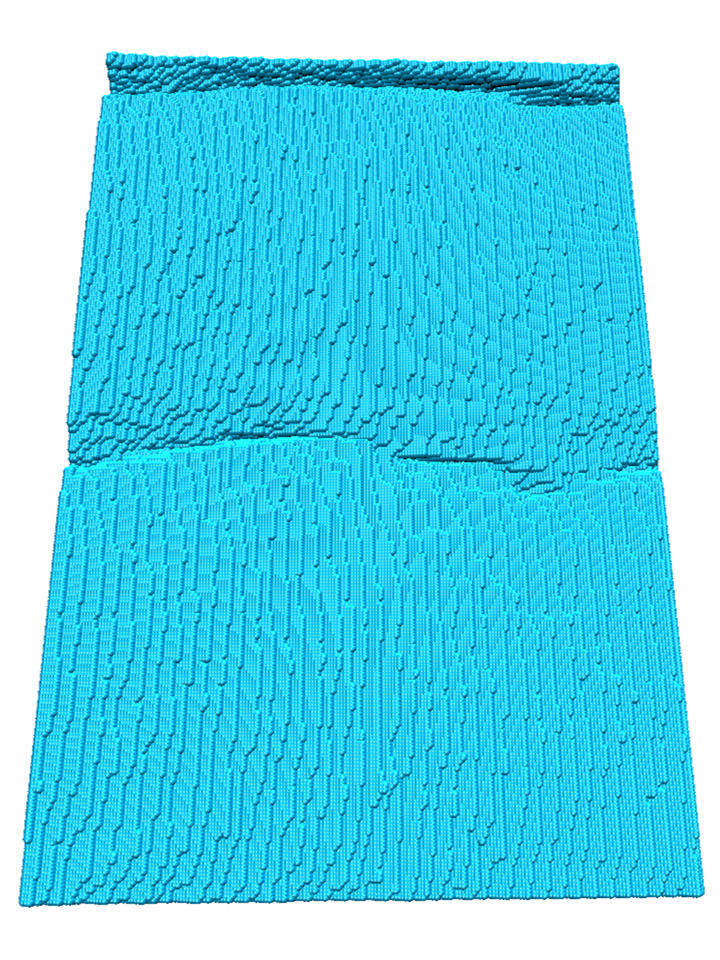}
c)\includegraphics[width=0.22\textwidth]{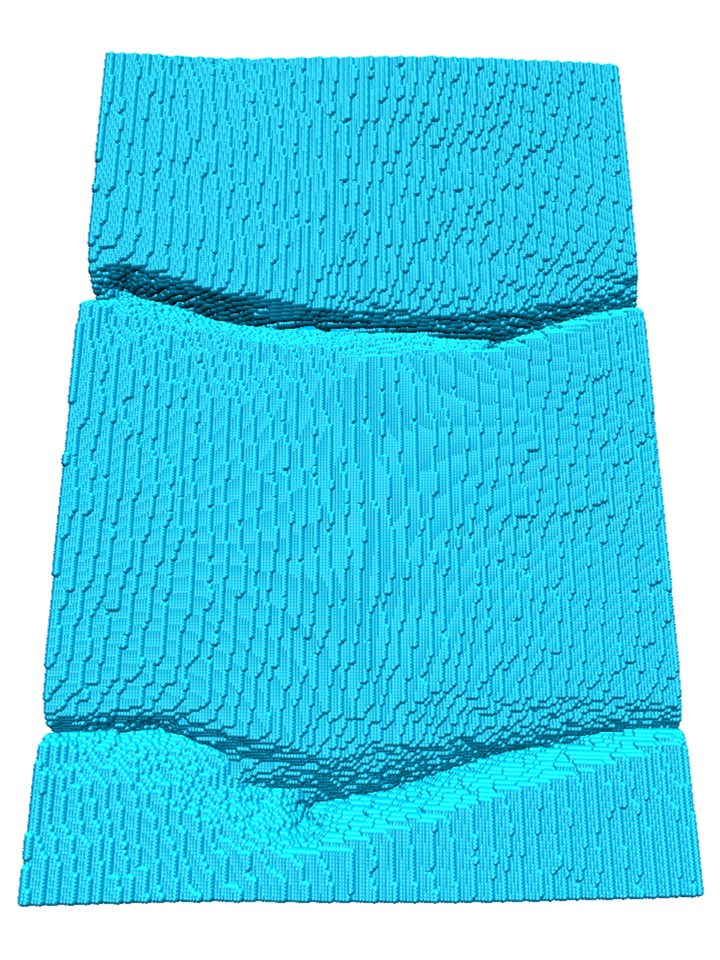}
d)\includegraphics[width=0.22\textwidth]{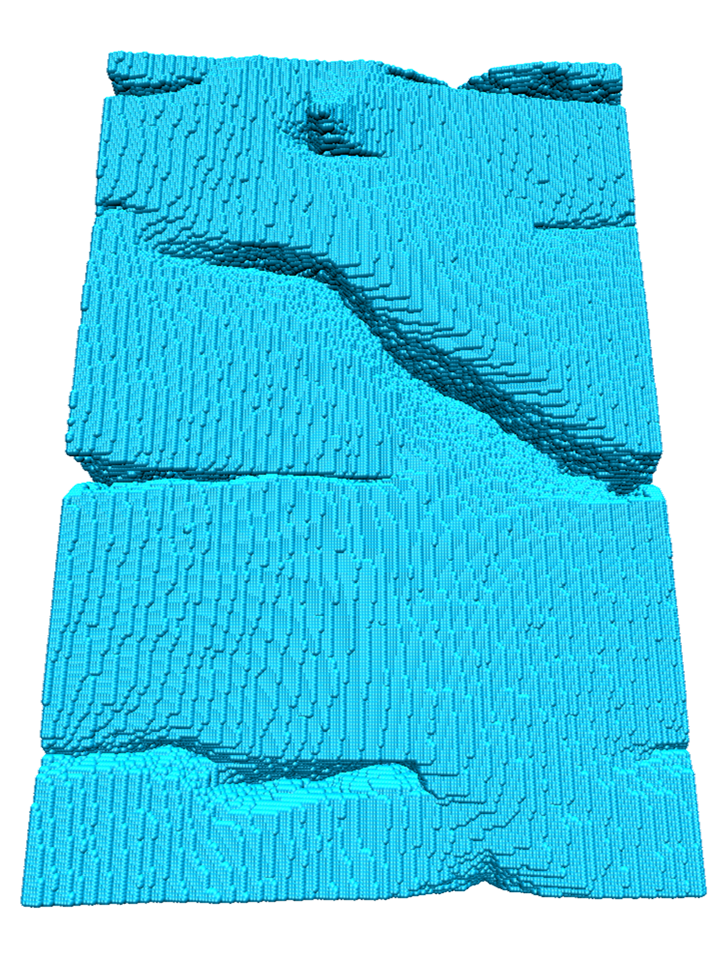}
\caption{Meanders obtained for $c_0 = 0.005$, $l_0 = 5$, $n_{DS} = 5$, $\beta E_V = 0.0$ and a) $\beta E_{ES} = 2.0$, b) $\beta E_{ES} = 4.0$, c) $\beta E_{ES} = 6.0$, d) $\beta E_{ES} = 8.0$, simulation time $t = 2 \cdot 10^6$. System size $200$ x $300$. }
\label{fig:pot_SB-dep}
\end{figure*}

The formation of meanders is typically attributed to the existence of an Ehrlich -- Schwoebel barrier at the top of the step. Consequently, we proceeded to assess the influence of this barrier as a subsequent step of our investigation. The ES height we examined within the range of $0.0$ to $10$, which corresponds to energy in the range of $0.0$~eV to $0.6$~eV, assuming a growth temperature of around $700~K$. The probability of jump across the step is in the range from $1$ to $4.5 \cdot 10^{-5}$. The results obtained for values equal to $\beta E_{ES}$: $2.0$, $4.0$, $6.0$ and $8.0$ in the absence of the potential well are presented in Figure~\ref{fig:pot_SB-dep}.
The introduction of the Schwoebel barrier into the system also results in the formation of meandering steps. However, the morphology of these meanders differs significantly from those observed in the presence of the potential well. The wavelength is considerably longer, and the curvature is significantly smaller. Additionally, the formed canyons between meanders are observed to be deeper. As in the previous case, increasing the height of the ES barrier leads to a shortening of the wavelength.

%%%%%%%%%%%%%%%%%%%%%%%%%%%%%%%%%%%%%%%%%%%%%%%%%%%%%%%%%%
%%%%%        Potential well and ES together          %%%%%
%%%%%%%%%%%%%%%%%%%%%%%%%%%%%%%%%%%%%%%%%%%%%%%%%%%%%%%%%%
\subsection{Mutual correlation between the potential well and the Ehrlich -- Schwoebel barrier}
\begin{figure*}[hbt]
 \centering
a)\includegraphics[width=0.22\textwidth]{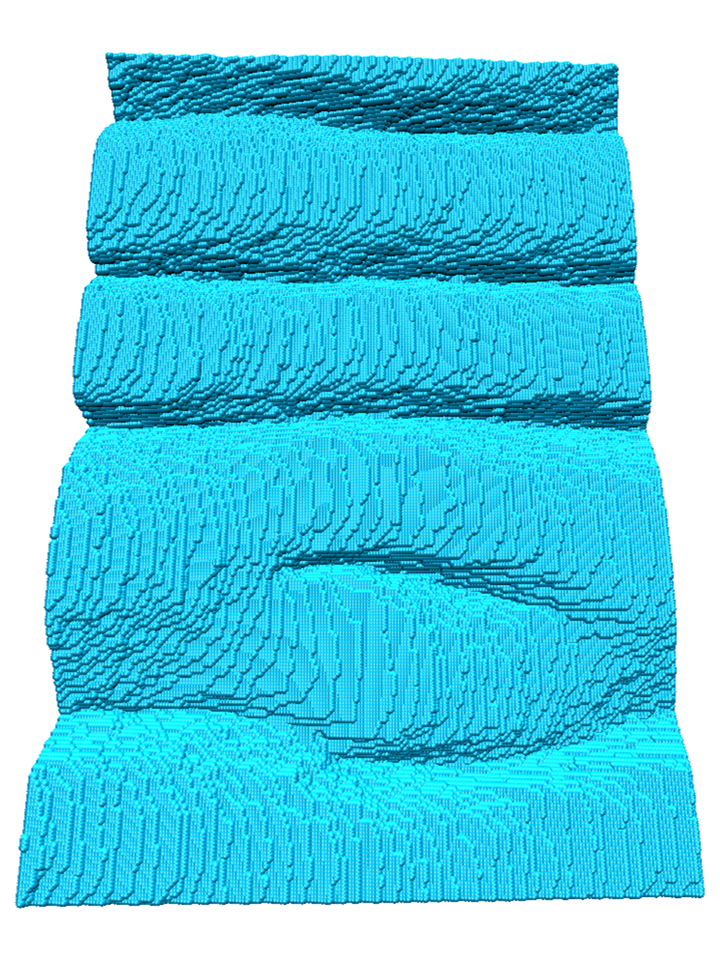}
b)\includegraphics[width=0.22\textwidth]{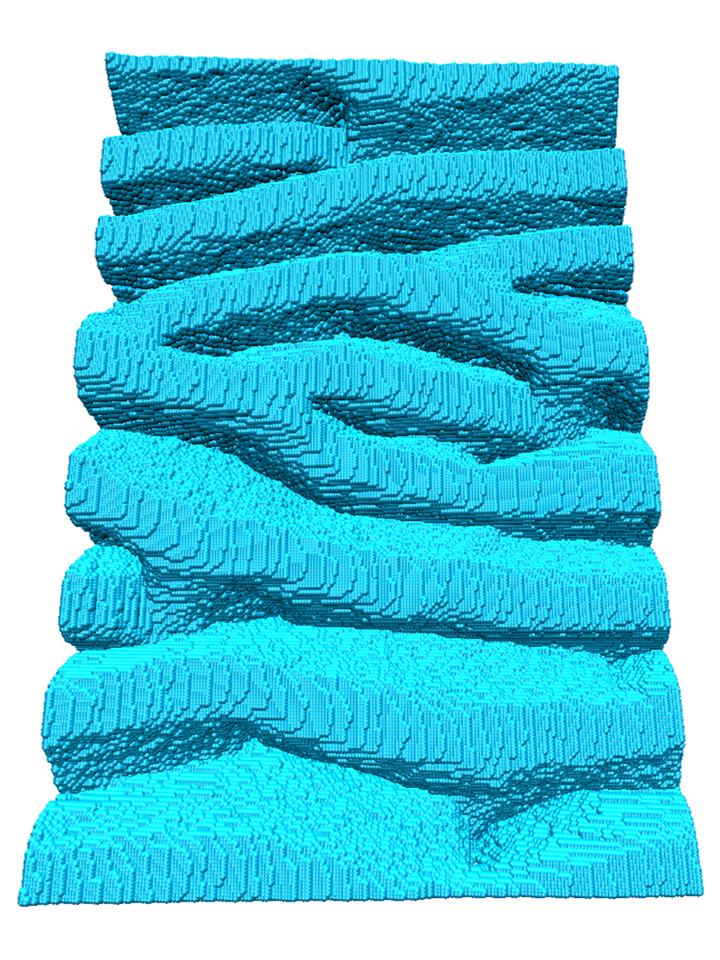}
c)\includegraphics[width=0.22\textwidth]{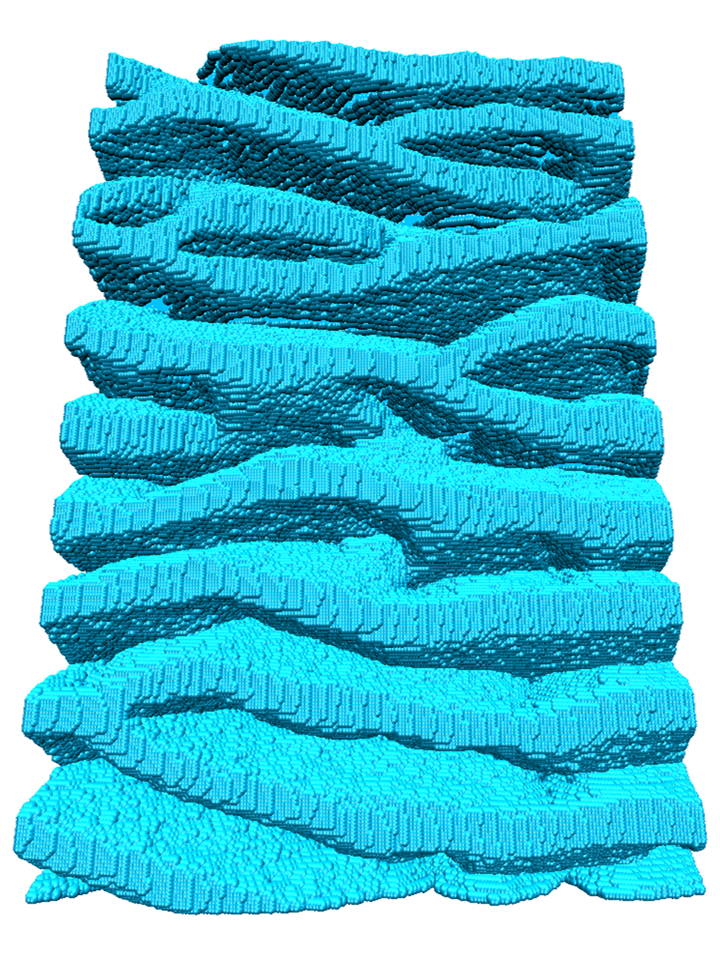}
d)\includegraphics[width=0.22\textwidth]{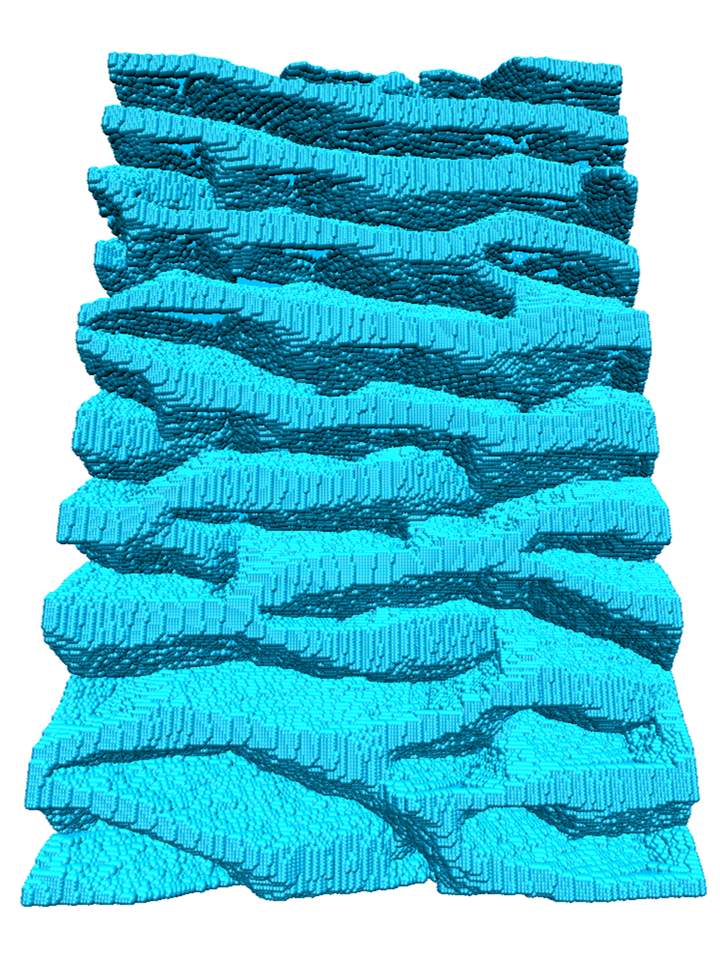}
\caption{Meanders obtained for $c_0 = 0.003$, $l_0 = 5$, $n_{DS} = 5$, $\beta E_V = 2.0$ and a) $\beta E_{ES} = 1.0$, b) $\beta E_{ES} = 3.0$,  c) $\beta E_{ES} = 5.0$, d) $\beta E_{ES} = 8.5$, simulation time $t = 2 \cdot 10^6$. System size $200$ x $300$. }
\label{fig:EB-dep}
\end{figure*}

The present study demonstrates that the mere presence of a potential well at the bottom of the step is sufficient to cause the formation of meanders  at the surface. Additionally, it has been verified that the ES barrier also results in meandering of the step, although the final shape differs. The following step involved an examination of the mutual correlation between the potential well and the Ehrlich -- Schwoebel barrier. Now the probability of jump across the step and out of the well is in the range from $1$ to $2 \cdot 10^{-9}$. The initial surface potential is now of the form illustrated in Figure~\ref{fig:potential}c. As expected, the introduction of the Ehrlich -- Schwoebel barrier into a system containing an existing potential well resulted in a significant modification of the final patterns, with a notable intensification of the meanders, as illustrated in  Figure~\ref{fig:EB-dep}.
An increase in the height of the ES results in a reduction in wavelength, and simultaneously to the formation of more meandering final structures. The presence of the ES also induces meandering of steps for the lower values  of potential well. The additional barrier causes the canyons formed between the meanders to become deeper than without the barrier, but when $\beta E_V$ is very large the canyons become shallower again. Additionally, as the depth of the potential well increases, the final structures become less ordered resembling surface roughening.

\begin{figure*}[hbt]
 \centering
a)\includegraphics[width=0.22\textwidth]{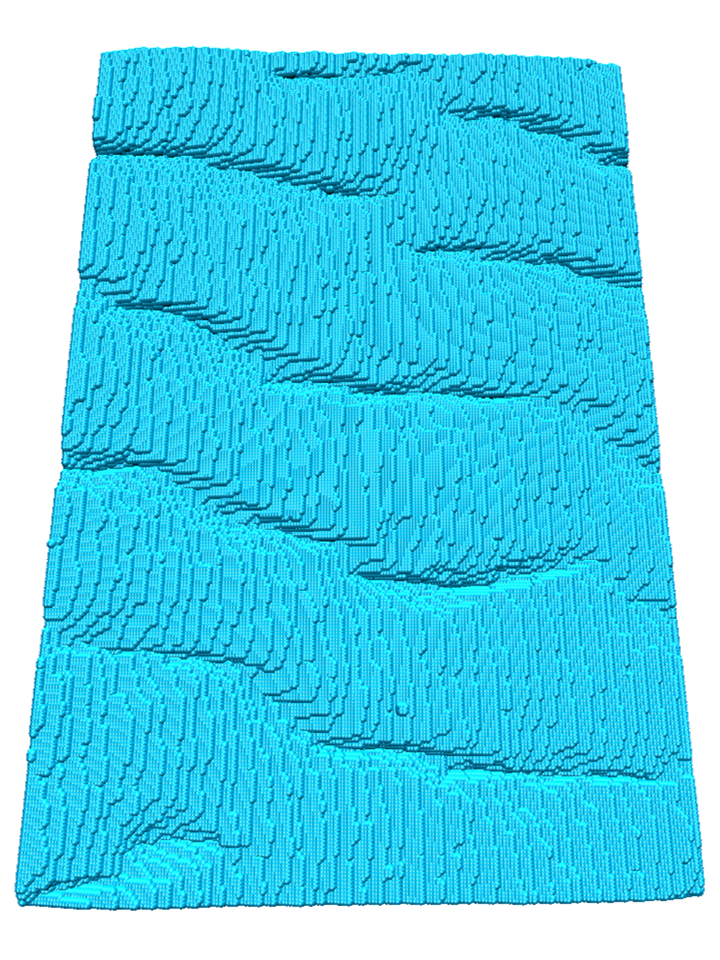}
b)\includegraphics[width=0.22\textwidth]{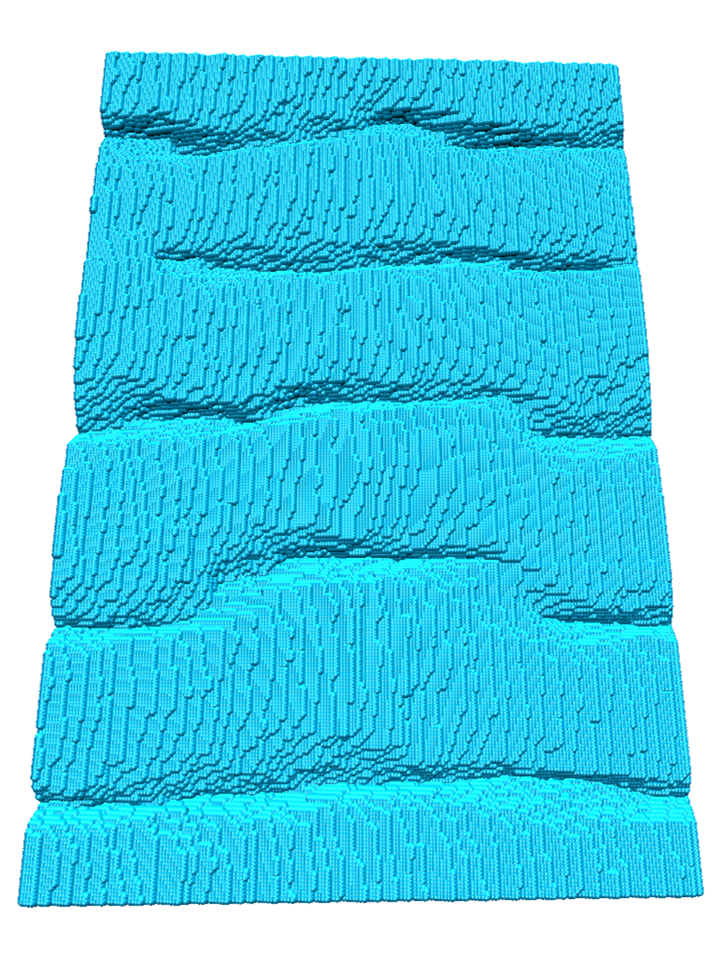}
c)\includegraphics[width=0.22\textwidth]{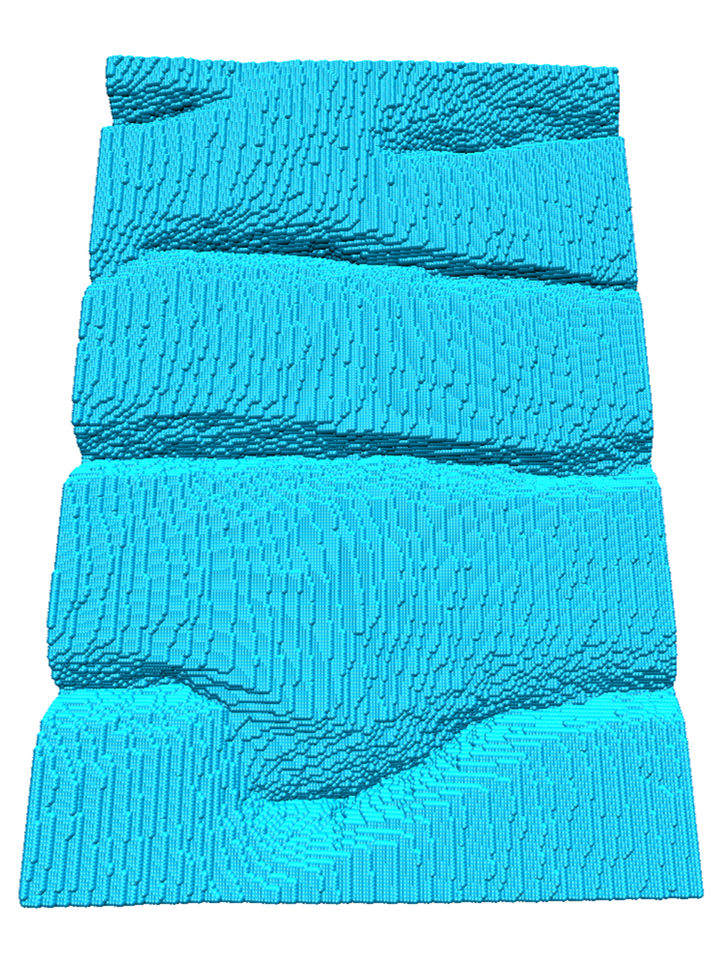}
d)\includegraphics[width=0.22\textwidth]{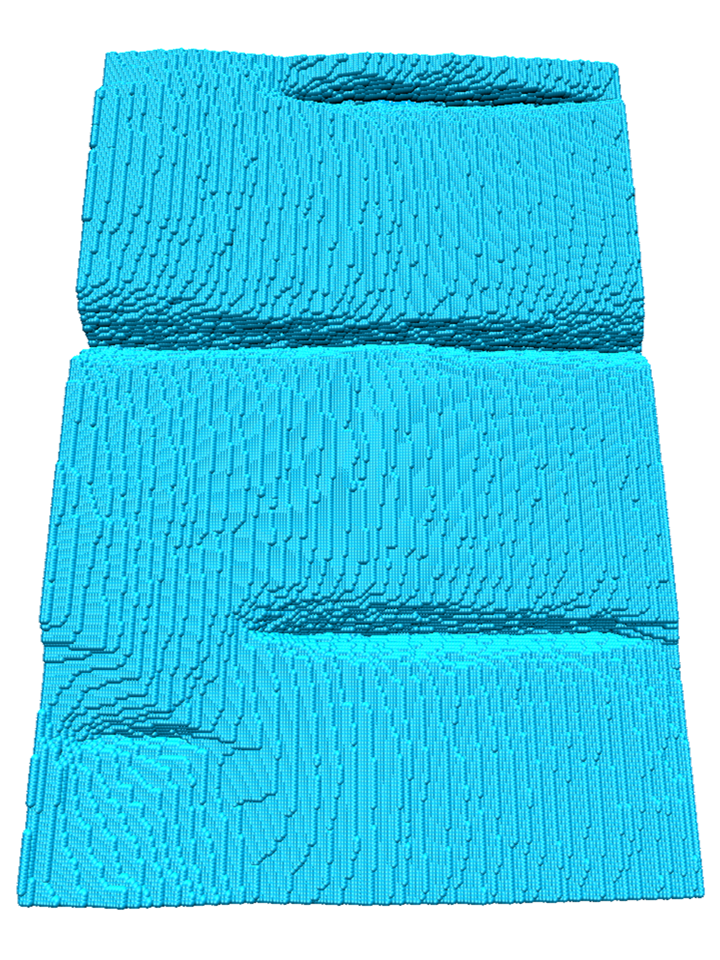}
\caption{Meanders obtained for $c_0 = 0.003$, $l_0 = 5$, $\beta E_V = 1.0$, $\beta E_{ES} = 3.0$ and a) $n_{DS} = 1$, b) $n_{DS} = 2$, c) $n_{DS} = 5$, d) $n_{DS} = 10$, simulation time $t = 2 \cdot 10^6$. System size $200$ x $300$. }
\label{fig:nds-dep}
\end{figure*}

%%%%%%%%%%%%%%%%%%%%%%%%%%%%%%%%%%%%%%%%%%%%%%%%%%%%%%%%%%
%%%%%%%%%%%%                nds                  %%%%%%%%%
%%%%%%%%%%%%%%%%%%%%%%%%%%%%%%%%%%%%%%%%%%%%%%%%%%%%%%%%%%
On the other side, meandered structure can also be influenced by the diffusion rate which in our model we can control by adjusting the number of diffusion jumps $n_{DS}$. So as a further step, we examined the dependence of the meanders on the diffusion rate and the obtained results are shown in the Figure~\ref{fig:nds-dep}. A comparison of the morphologies obtained for different values of $n_{DS}$ clearly shows that increasing the diffusion rate leads to an increase in the wavelength of the meander.

%%%%%%%%%%%%%%%%%%%%%%%%%%%%%%%%%%%%%%%%%%%%%%%%%%%%%%%%%%
%%%%%              Akiyama potentials                %%%%%
%%%%%%%%%%%%%%%%%%%%%%%%%%%%%%%%%%%%%%%%%%%%%%%%%%%%%%%%%%
\subsection{Realistic  potentials}
To validate our model, we decided to use in the simulations the realistic surface potentials which simultaneously could be easily incorporated into our model. To achieve this, we used the energy potential derived from \textit{ab-initio} calculations performed by T. Akiyama and co-workers (Ref.~\cite{Akiyama-JCG21}). The energies obtained through density functional theory (DFT) calculations are determined for the system at 0K, resulting in relatively high energy barrier values. At higher temperatures typical of the crystal growth process, where the entire crystal vibrates, adatoms likely encounter effectively lower energy barriers. These reduced barriers can be overcome via thermally activated processes. Rather than relying on the exact numerical values of the calculated energies, we focused on their relative relationships. Specifically, our attention was directed toward the most prominent features of the potential landscape: the potential wells and the Ehrlich -– Schwoebel barrier, as identified in ~\cite{Akiyama-JCG21}.

\begin{figure}[hbt]
 \centering
a)\includegraphics[width=0.22\textwidth]{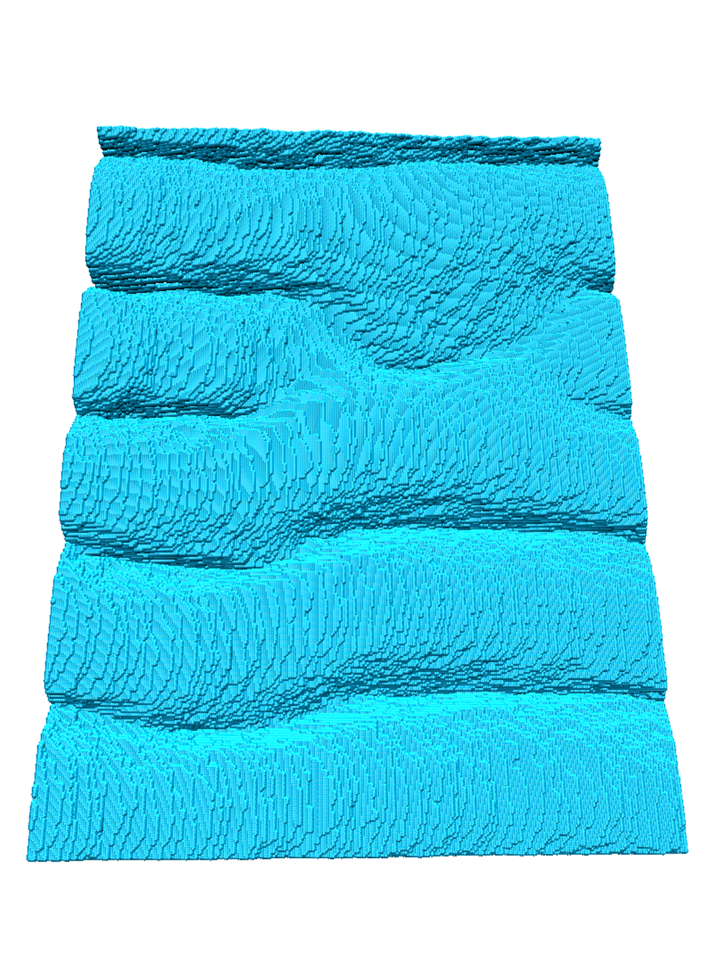}
b)\includegraphics[width=0.22\textwidth]{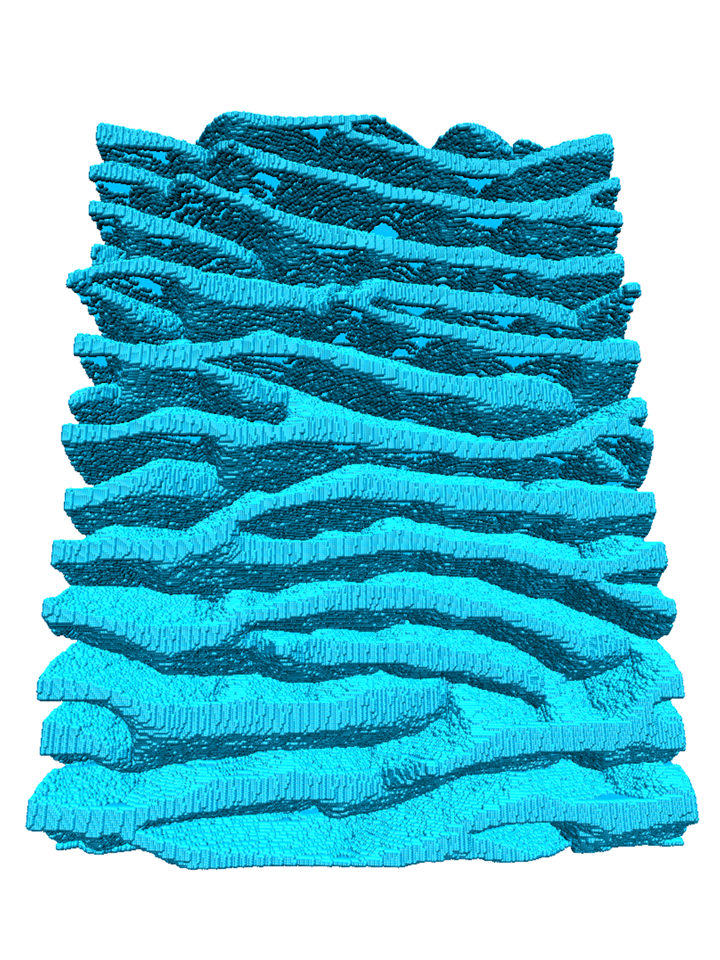}
c)\includegraphics[width=0.22\textwidth]{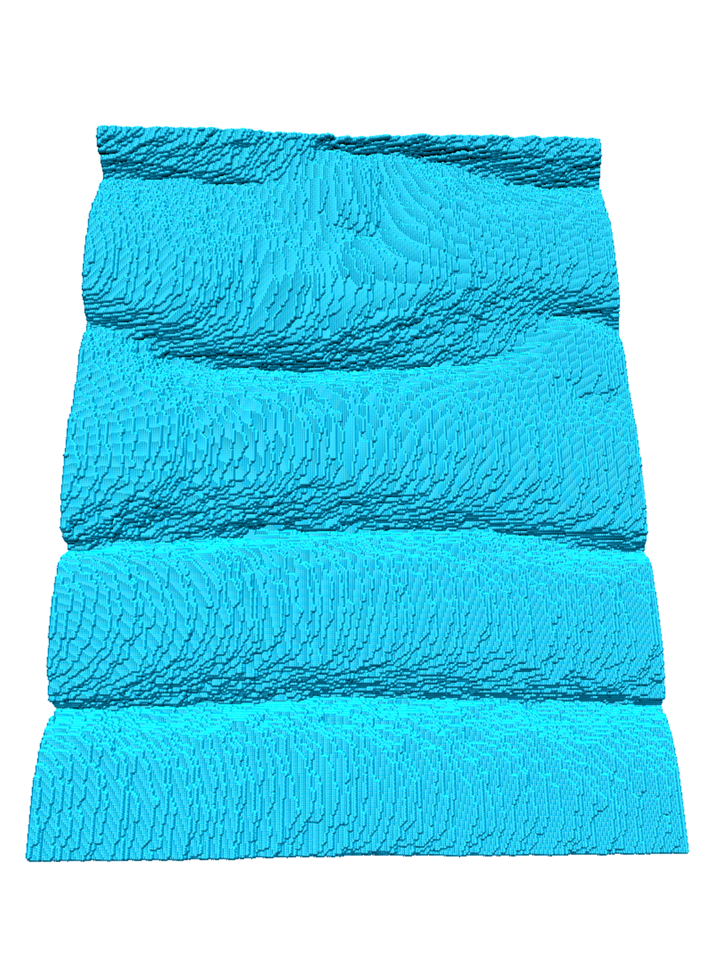}
d)\includegraphics[width=0.22\textwidth]{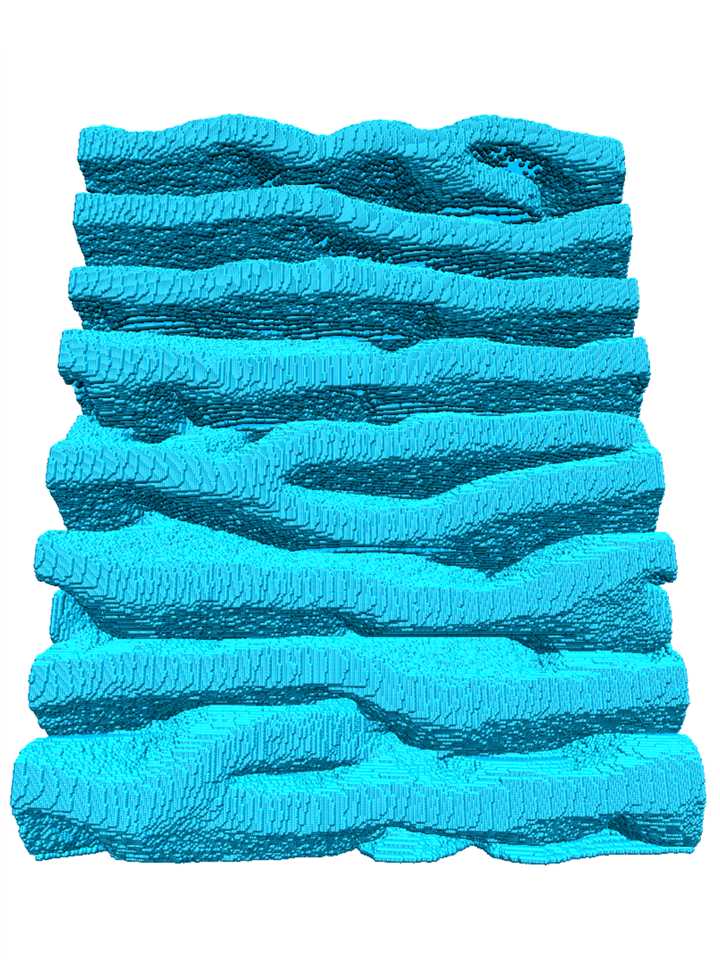}
\caption{Meanders obtained for $c_0 = 0.003$, $l_0 = 5$, $n_{DS} = 5$, $\beta E_V = 2.0$ and a) $\beta E_{ES} = 0.62$, b) $\beta E_{ES} = 7.32$, c) $\beta E_{ES} = 0.54$, d) $\beta E_{ES} = 3.42$, simulation time $t = 2 \cdot 10^6$. System size $300$ x $400$. Relations between values of $\beta E_V$ and $\beta E_{ES}$ are taken from Ref. \cite{Akiyama-JCG21}. }
\label{fig:Aki-dep}
\end{figure}

By adapting the aforementioned landscape to our model, we were able to determine the relations between the energy barriers, specifically how high the Ehrlich –- Schwoebel barrier is in comparison to the value of the potential well. This procedure leads to the following dependencies: $\beta E_{ES}\approx 0.31 \beta E_{V}$, $\beta E_{ES}\approx 3.67 \beta E_{V}$, $\beta E_{ES}\approx 0.27 \beta E_{V}$ and $\beta E_{ES}\approx 1.67 \beta E_{V}$. These values are taken from Figures 3a, 3b, 4a, and 4b in Ref.~\cite{Akiyama-JCG21}, respectively. Assuming a potential well depth of $\beta E_{V} = 2.0$, the height of the ES barrier is equal to $\beta E_{ES} = 0.62, 7.32, 0.54$ and $3.42$, respectively. The resulting morphologies from our simulations with these values are illustrated in Figure~\ref{fig:Aki-dep}. In experiments, during the epitaxial growth of GaN by metal–organic vapor phase epitaxy (MOVPE), meanders are observed at temperature 1150~K \cite{Pandey-Vac}. At this temperature, assumed above value of the potential well corresponds to the energy equal to $0.25$~eV and values of the Ehrlich -- Schwoebel barrier to energies equal to $0.08$~eV, $0.9$~eV, $0.07$~eV, $0.42$~eV, respectively.
It can be seen that the presented results reproduce qualitatively the meanders observed experimentally  \cite{Damilano-JAP,Chou-ASS,Wu-JCG,Turski-ASS,Gocalinska-ASS,Pandey-Vac}.

\subsection{Diagrams of meander wavelength and pattern formation}
Further, to analyze quantitatively the obtained meandered structures, we measure the wavelength of meanders by means of height--height correlation function $C(r) = C(r_i - r_j) = \langle [h(r_i) - h(r_j)]^2 \rangle $ which describes the average height difference between any two points $i$ and $j$ on the surface separated by distance $r = r_i - r_j$. In addition, the correlation function can be calculated along x and y axis separately, and in this study we are interested in its calculation along y axis which is the direction along steps in which the meanders develop. Thus, in the case of a regular and ordered meander structure, this correlation function exhibits oscillatory behavior, repeating itself regularly at a distance equal to the meander wavelength.  Such behavior we observe in our meandered surface morphologies and their characteristic wavelength is measured as the position of the first minimum of the corresponding correlation function.

\begin{figure*}[hbt]
 \centering
a)\includegraphics[width=0.45\textwidth]{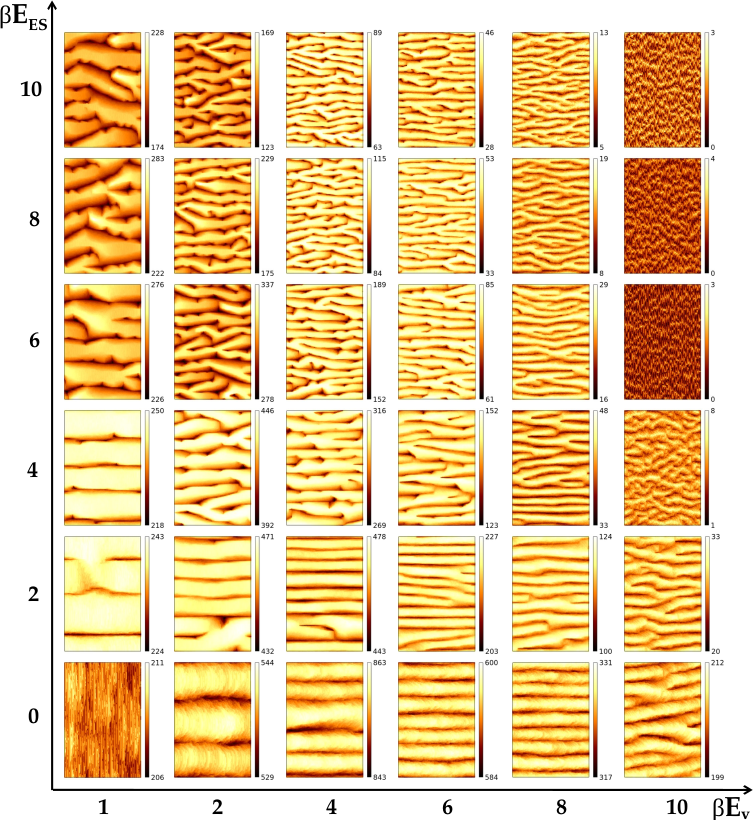}
\hspace{0.1cm}
b)\includegraphics[width=0.5\textwidth]{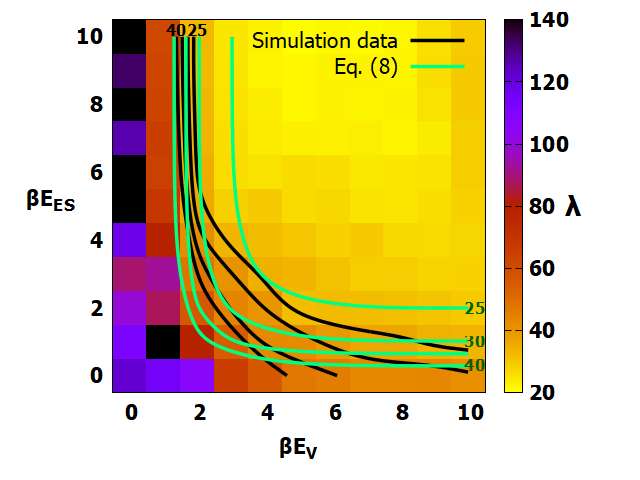}
\caption{a) Diagram of pattern formation on the surface obtained for  $c_0 = 0.003$, $l_0 = 5$, $n_{DS} = 5$ and $t = 2 \cdot 10^6$ at different values of potential well and Schwoebel barrier. Different colors correspond to different surface height. b) Diagram of meander wavelength $\lambda$ as a function of the depth of the potential well $\beta E_{V}$ and the height of the Ehrlich -- Schwoebel barrier $\beta E_{ES}$ obtained for  $c_0 = 0.003$, $l_0 = 5$, $n_{DS} = 5$ and $t = 2 \cdot 10^6$. Different colors correspond to different meander wavelength. The presented isolines correspond to $\lambda = 25, 30, 35, 40$ and are extracted from simulation data and Eq.~(\ref{lambda_fit}) respectively.}
\label{fig:diagram}
\end{figure*}

Based on this analysis, we perform a systematic study of the meander wavelength $\lambda$ measured quantitatively for morphologies resulting at different values of the potential well and Schwoebel barrier and investigate in detail the mutual correlation between both of them. The observed surface morphologies can be presented in a proper diagram of pattern formation as shown in Figure~\ref{fig:diagram}a. The results for $\lambda$ obtained from our simulation data and averaged over 10 simulation runs are presented in Figure~\ref{fig:diagram}b. This colored diagram, illustrating the dependence of the meander wavelengths as a function of the depth of the potential well and the height of the Schwoebel barrier, appears as a very useful and informative to determine the potential regions where meanders of proper wavelength can develop. As one can see, at zero potential well and zero Schwoebel barrier meanders do  not emerge. For non-zero barriers, the steps start to meander, leading to increasingly shorter wavelengths as the depth  of potential well increases. The same behavior can be observed if the Schwoebel barrier becomes higher. The combination of both energies causes an even more dramatic reduction in the meander wavelength. The presented on this diagram isolines, corresponding to values of $\lambda$ between $25$ and $40$, define the potential region where well-formed meanders can be obtained. In the bright region above the isoline $\lambda = 25$, when both energies (potential well and Schwoebel barrier) are too high, regular meanders cannot develop; only an early stage of meandering (or no meandering) is observed and the surface remains rough.

%%%%%%%%%%%%%%%%%%%%%%%%%%%%%%%%%%%%%%%%%%%%%%%%%%%%%%%%%%
%%%%%          Formation of the meanders             %%%%%
%%%%%%%%%%%%%%%%%%%%%%%%%%%%%%%%%%%%%%%%%%%%%%%%%%%%%%%%%%
\subsection{The mechanism of meander formation}
\begin{figure*}[hbt]
 \centering
a)\includegraphics[width=0.3\textwidth]{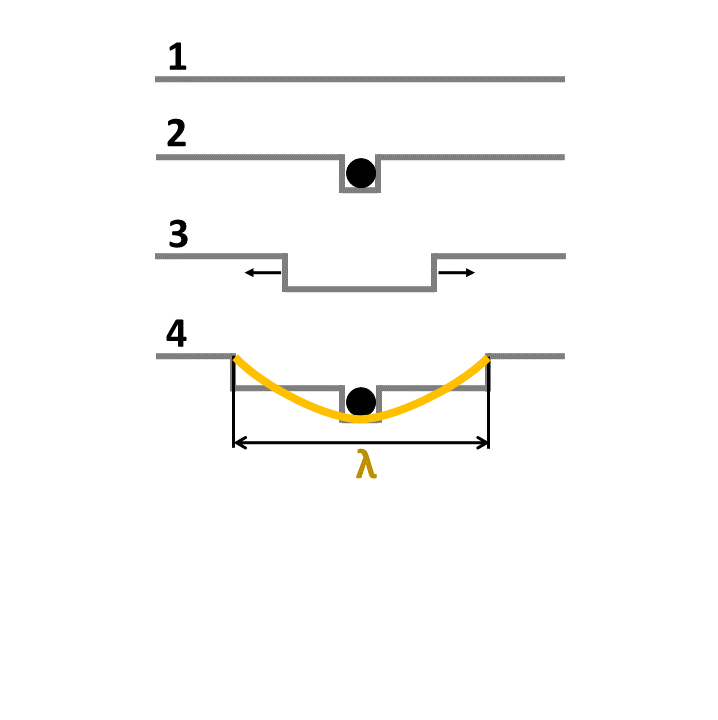}
b)\includegraphics[width=0.3\textwidth]{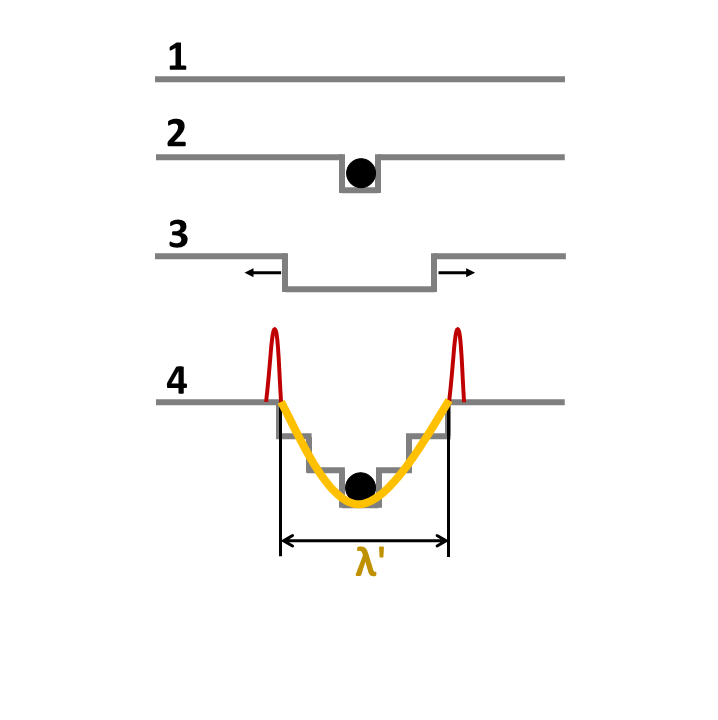}
c)\includegraphics[width=0.25\textwidth]{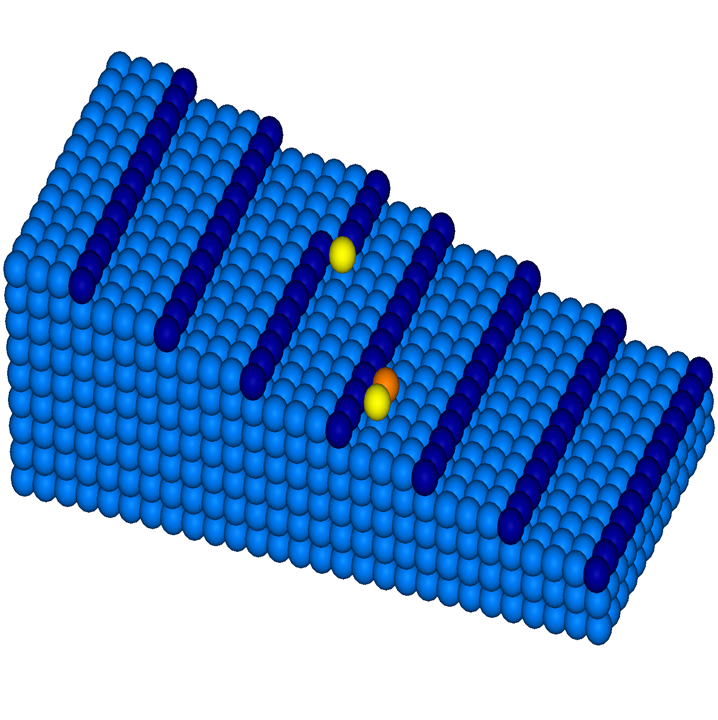}
\caption{Visualization of the mechanism of the formation of the meanders a) without, b) with the Ehrlich –- Schwoebel barrier present. c) Initial vicinal surface with the adatoms (yellow ball) which become part of the crystal and are involved in the formation of the meanders.}
\label{fig:formation}
\end{figure*}

To elucidate the mechanism of meander formation, it is clear that kinks play a pivotal role~\cite{Kallunki-PRB,Hamouda-SS}. The vicinal cellular automaton model allows us to influence the concentration of kinks by controlling the process of their creation. This process begins with an initially straight step (Figure~\ref{fig:formation}a-1). An adatom diffusing on the surface then reaches the step and attaches to it with a certain probability, denoted as $P_s$ (Figure~\ref{fig:formation}a-2). This attachment automatically generates two kinks that migrate in opposite directions (Figure~\ref{fig:formation}a-3) due to the adsorption of adatoms at the kinks with probability $P_k$. In the model, we assume that an adatom reaching a kink is always incorporated into the crystal. As a result, both kinks move with a velocity equal to $aP_k$, where $a$ is the lattice constant and $P_k$ measures attachment probability per MC time unit.
Then, another diffusing adatom reaches the newly formed straight step and attaches to it (Figure~\ref{fig:formation}a-4) with  probability $P_s$. Thus,  approximately  time between  first  and  the  second  attachment to the step is equal to $P_s^{-1}$. This means that the meander starts to form after the kinks have moved along the distance
\begin{equation}
\lambda \approx \frac{aP_k}{P_s},
\end{equation}
that is approximately equal to a meander wavelength $\lambda$. Note, that we can estimate the wavelength of meander in such a way, but we do not prejudge if meanders would appear at all.

The presence of a Schwoebel barrier on the top of the step results in diffusional screening, which prevents adatoms from reaching the kinks (Figure~\ref{fig:formation}b-4).
In this case, the step movement gets faster in the central part  of meander  and rather slower away from this region due to the limited number of adatoms  arriving at these distant locations. As a consequence, the resulting meanders are narrower and exhibit greater curvature, whereas large and deep canyons are formed between them. Such mechanism of meander formation is observed in the presence of Schwoebel barrier, as demonstrated in Figures~\ref{fig:pot_SB-dep} and \ref{fig:EB-dep} for high values of $\beta E_{ES}$.

Let us estimate the probability of incorporation. As mentioned earlier, an adatom is incorporated into the crystal at step only if it has another adatom in its immediate vicinity, as illustrated in Figure~\ref{fig:formation}c. Thus, we can calculate the probability $P_s$ as equal to $\rho_s^2$. We approximate the density $\rho_s$, defined as a local particle density at step, by its equilibrium value, assuming that attraction to the step is so slow, that allows for establishing equilibrium. Thus, we have

\begin{equation}
(e^{-\beta E_{ES}}+1) \rho_l+2 \rho_s=(e^{-\beta E_{ES}}e^{-\beta E_{V}}+e^{-\beta E_{V}}+2+\frac{\rho_s}{n_{DS}}) \rho_s
\end{equation}
what records balance of streams coming in and out the site at step and where $\rho_l$ is a local particle density  at terrace. To complete this equation, an expression describing the relationship between $\rho_l$, $\rho_s$, and $c_0$ should be introduced. However, since the local density at the terrace is not in equilibrium — adatoms are continuously adsorbed and delivered — the process remains quasi-stationary. Consequently, the exact form of this condition is not precisely known. Instead, we adopt a general relation:
\begin{equation}
   \rho_l \alpha+\rho_s= \gamma c_0
\end{equation}
denoting the balance between the step surroundings and certain terrace areas. The variables $\alpha$ and $\gamma$ may depend on external particle flux and the shape of the potential. However, within specific parameter ranges, their variation is expected to be minimal.
To simplify, we approximate these variables by their mean values and treat them as functions of all parameters of the model other than the potential, allowing for adjustment as needed. Combining all these considerations gives:

\begin{equation}
\rho_s=\frac{\gamma c_0}{e^{-\beta E_V}\alpha+1}
\end{equation}
in the first term of expansion in small $c_0$. Now, at kink sites, where the local particle density is defined as $\rho_k$, we expect far from equilibrium situation, because any adatom that arrives to kink is  immediately attached to the crystal. Thus, to calculate probability $P_k$ as equal to $\rho_k$, we sum up  streams of adatoms from all four directions
\begin{equation}
    \rho_k=0.25(\rho_s+\rho_l+2e^{-\beta E_{ES}}\rho_l).
\end{equation}
Finally, after approximating that in this case $\rho_s \approx \rho_l$, and fitting the formula to extensive data obtained for various model parameters, we determine the values of  $\alpha$ and $\gamma$ as  functions  of $n_{DS}$ and  $l_0$, yielding the expression:
\begin{equation}
 \lambda = \frac{\rho_k}{\rho_s^2} = \frac{0.066}{c_0} (1 + e^{-\beta E_{ES}})(0.25 \sqrt{n_{DS}} l_0 e^{-\beta E_{V}} + 1 ).
\label{lambda_fit}
\end{equation}
Contours for selected values of the meander wavelength are presented in Figure~\ref{fig:diagram}b. The derived expression provides an approximation that effectively captures the dependence of $\lambda$ on $E_V$ and $E_{ES}$, as shown in Figure~\ref{fig:diagram}b, even though it does not fully replicate the exact relationships observed in the simulation. The presence of terms with exponents featuring different variations of $E_V$ and $E_{ES}$ highlights the independent and combined effects of the potential well and the Schwoebel barrier on the wavelength $\lambda$, while the prefactors reflect the influence of various model parameters, such as $c_0$, $l_0$, and $n_{DS}$.
A more rigorous and detailed analysis will be necessary to gain a deeper understanding of the complexities of the meandering process, which will remain a primary focus of our future research.

%%%%%%%%%%%%%%%%%%%%%%%%%%%%%%%%%%%%%%%%%%%%%%%%%%%%%%%%%%%%%%%%%%%%%%%%
%                              Conclusions
%%%%%%%%%%%%%%%%%%%%%%%%%%%%%%%%%%%%%%%%%%%%%%%%%%%%%%%%%%%%%%%%%%%%%%%%
\section{Conclusions}
In this study, we conduct a thorough investigation into the intricate relationship between meander formation and the surface potential energy landscape using the Cellular Automaton model. Our findings highlight the critical role that surface energy potential plays in shaping both the formation and evolution of meanders.
One of the most notable discoveries of our analysis is that meanders can form solely due to the presence of a potential well at the bottom of a step, without the need for additional complex conditions. By systematically examining how meander behavior varies with the depth of the potential well, represented by $\beta E_V$, we observed a clear trend: as the depth of the potential well increases, the wavelength of the meanders decreases. This observation provides deeper insights into the fundamental mechanisms driving meander formation.
Additionally, we examine the influence of the Ehrlich -- Schwoebel barrier on meander formation. Interestingly, our analysis reveals that while meanders can indeed form solely due to the Ehrlich -- Schwoebel barrier, their shape is markedly different from those formed by the potential well alone. The balance between the potential well and the Ehrlich -- Schwoebel barrier creates meanders with unique characteristics -- shorter wavelengths that decrease further as the Schwoebel barrier height increases, resulting in increasingly intricate and complex meander patterns.

To explain this behavior, we propose a mechanism for meander formation. This mechanism suggests that the length of the meanders is determined by the ratio of the probabilities of adatom attachment to kinks and straight steps. Both of these probabilities are functions of local adatom densities. The derived expression successfully captures the observed relationship between meander length and both the Ehrlich -- Schwoebel barrier height, $\beta E_{ES}$, and the potential well depth, $\beta E_V$. These findings not only enhance our understanding of surface pattern formation but also pave the way for further exploration into the dynamics of crystal growth and surface morphology.

The primary mechanism of pattern formation analyzed here is adparticle diffusion within the surface potential. The shape of the surface potential plays a decisive role in determining the final morphology of the surface. On the other hand, surface potential landscapes, as experienced by particles of different types diffusing along the surface, can be calculated using DFT methods. By understanding these landscapes, we can predict the possible behavior of the surface during the crystal growth process. This knowledge enables us to identify optimal conditions for the process, such as the appropriate temperature — which influences the overall diffusion rate — as well as the flux rate and other critical parameters.

\section*{Acknowledgments}
The authors thank to Vesselin Tonchev for valuable discussions. Part of the calculations were done on HPC facility Nestum (BG161PO003-1.2.05).
The authors thank The Polish National Center for Research and Development (grant no. EIG CONCERT-JAPAN/9/56/AtLv-AlGaN/2023),
The Bulgarian National Science Fund (grant No. KP-06-DO02/2/18.05.2023), the Polish Academy of Sciences
and the Bulgarian Academy of Sciences (grant No. IC-PL/07/2024-2025) for financial support.

\section {Supplementary Information}
\renewcommand{\thefigure}{\textbf{S\arabic{figure}}}
\setcounter{figure}{0}
In this supplement we provide  additional Figures that show the behavior of a single step (Fig.~\ref{fig:one-step}), the scaling of meanders with the system size (Fig.~\ref{fig:size}), their evolution over time (Fig.~\ref{fig:time}), and the changes in the surface pattern that occur when nucleation of islands on terraces is allowed or forbidden for two different terrace lengths and potential well depths (Fig~\ref{fig:nucleation}).

\begin{figure*}[hbt]
 \centering
a)\includegraphics[width=0.4\textwidth]{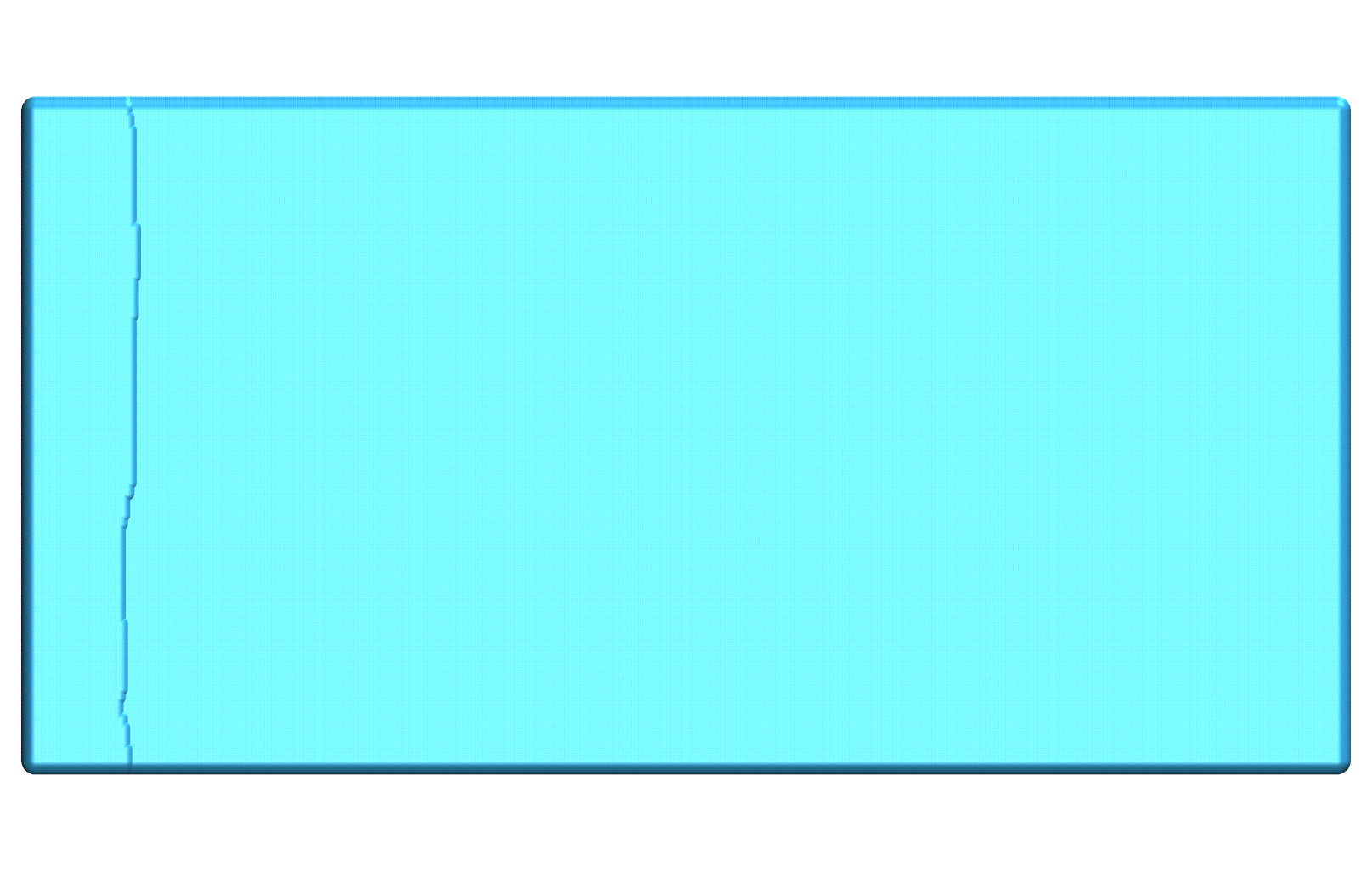}
b)\includegraphics[width=0.4\textwidth]{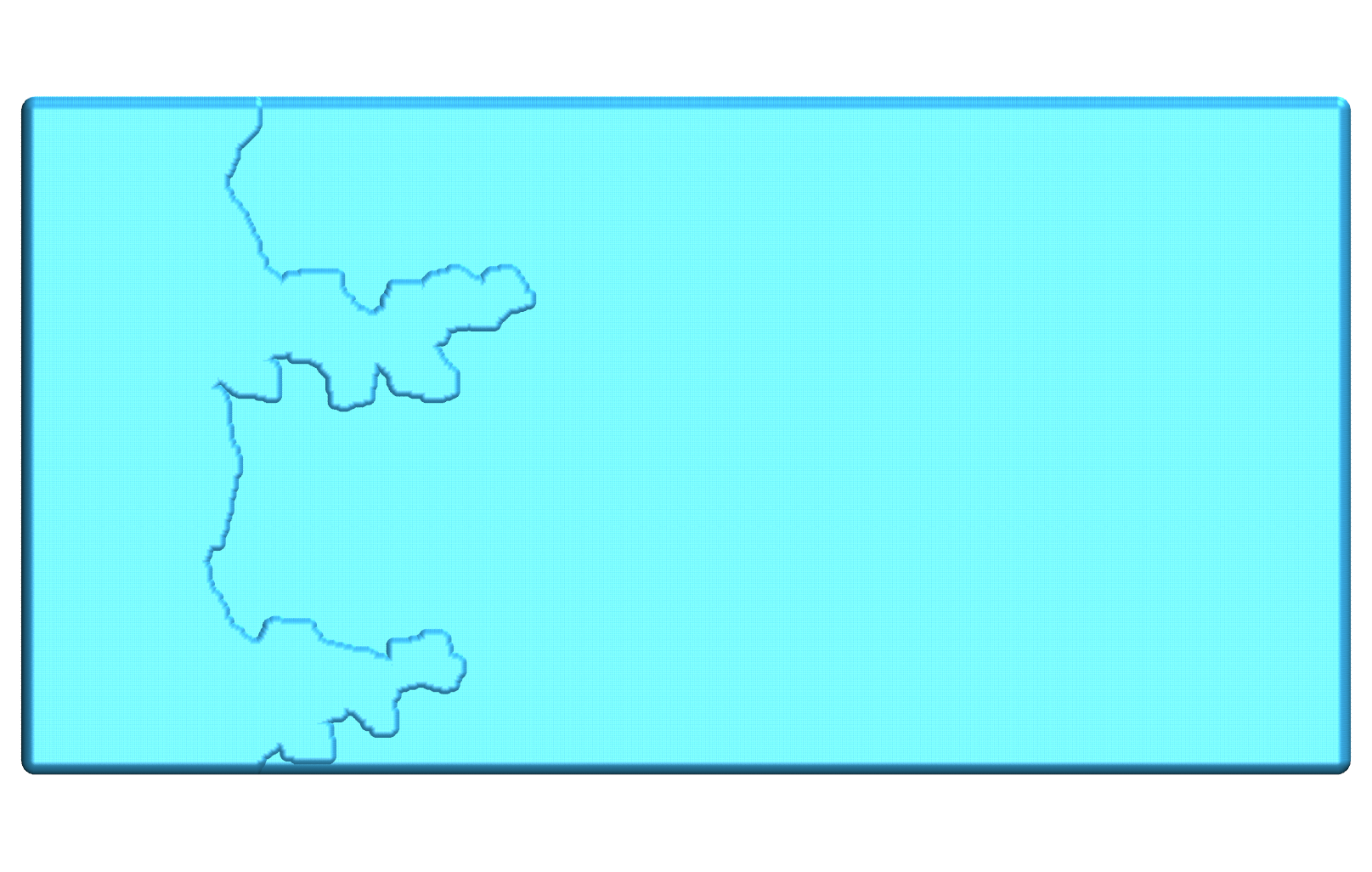}
\caption{\textbf{Single step evolution. a)} Straight step obtained for a system with a single step when the potential well and the Ehrlich-Schwoebel barrier are absent, \textbf{b)} Meandering step derived for a system when the potential well is present and equal to $\beta E_{V} = 6$, $\beta E_{ES} = 0$.  Simulations were performed for $c_0 = 0.005$, $l_0 = 600$, $n_{DS} = 3$, $t = 2*10^6$. System size $600$ x $300$.}
\label{fig:one-step}
\end{figure*}

\begin{figure*}[hbt]
 \centering
a)\includegraphics[width=0.25\textwidth]{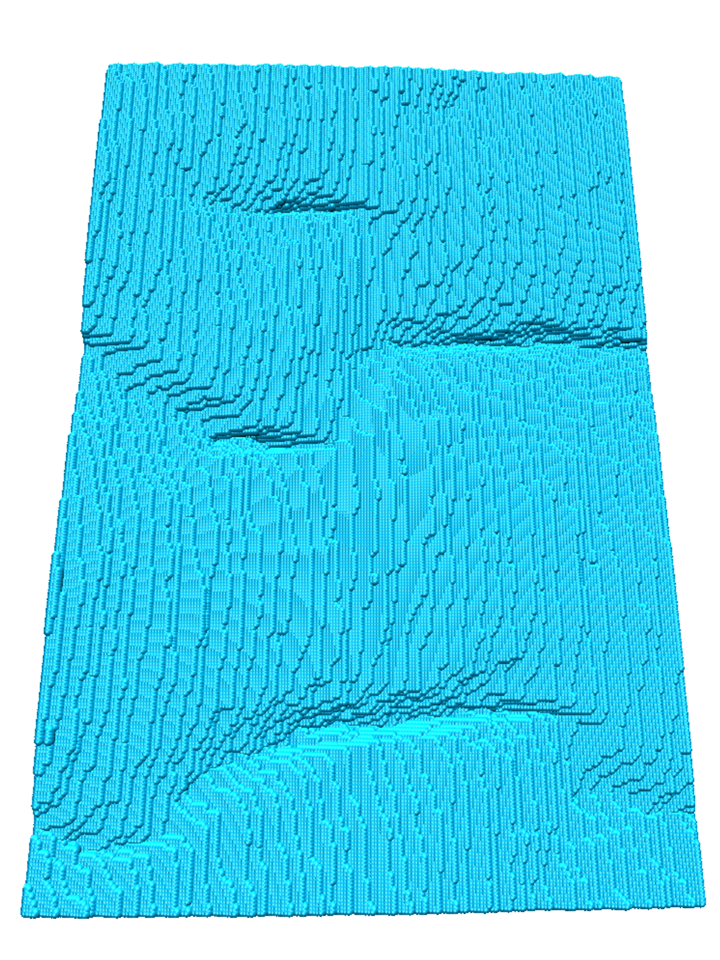}
b)\includegraphics[width=0.25\textwidth]{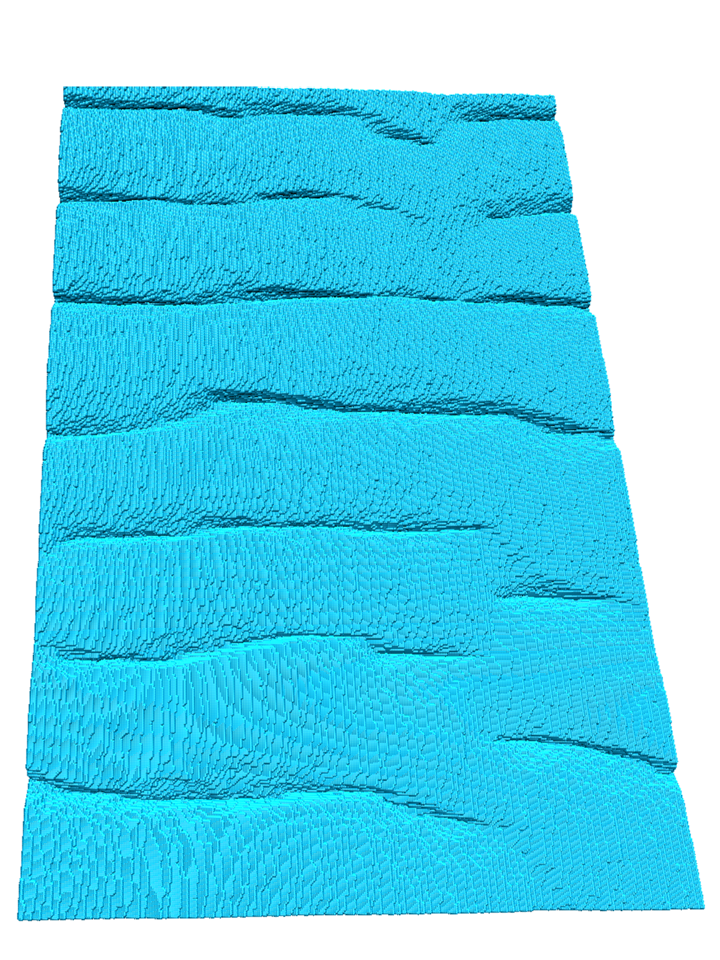}
\caption{\textbf{Different system sizes.} The scaling of the system was examined in relation to its dimensions, with meanders obtained for a system sizes of \textbf{a)} $200$ x $300$, and \textbf{b)} $400$ x $600$.
The calculations were performed for the following parameters: $c_0 = 0.003$, $l_0 = 5$, $\beta E_{V} = 1.0$, $\beta E_{ES} = 2.5$, $n_{DS} = 5$,  $t = 10^6$.}
\label{fig:size}
\end{figure*}

\begin{figure*}[hbt]
 \centering
a) \includegraphics[width=0.22\textwidth]{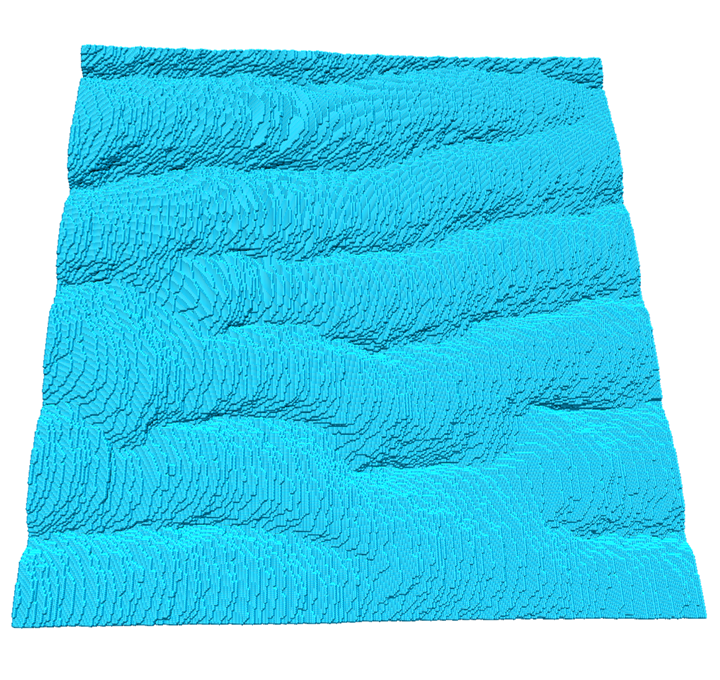}
b) \includegraphics[width=0.22\textwidth]{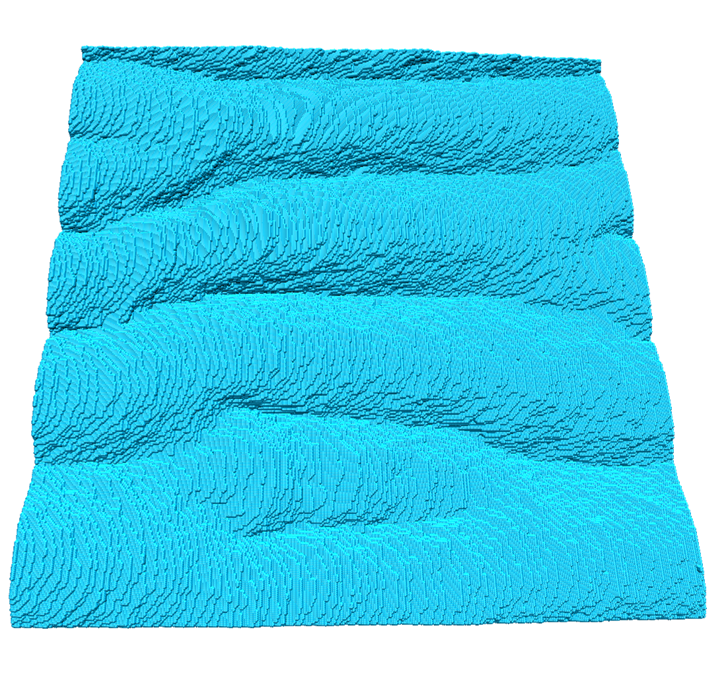}
c) \includegraphics[width=0.22\textwidth]{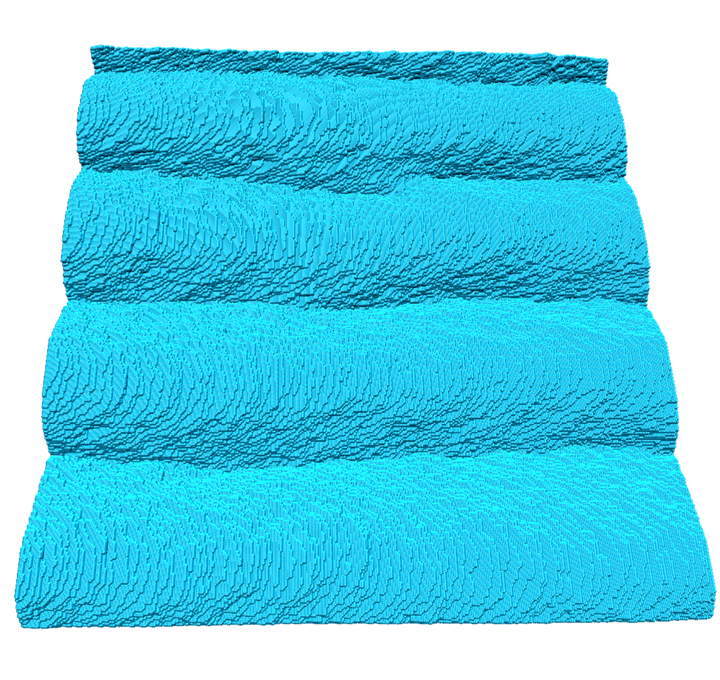}
d) \includegraphics[width=0.22\textwidth]{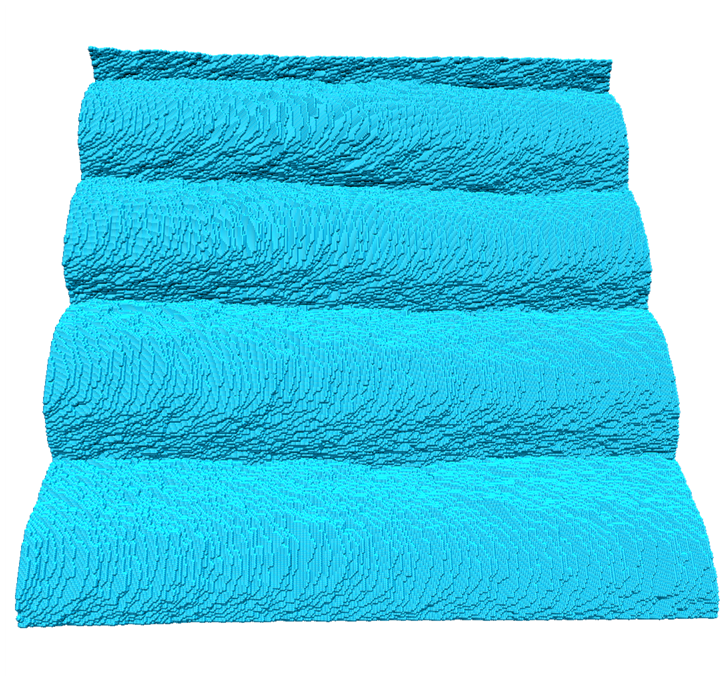}
\caption{\textbf{Time evolution.} Surface pattern with meanders showing the straightening of the meanders over time. Results are shown for times \textbf{a)} $t = 0.20 \cdot 10^7$, \textbf{b)} $t = 0.46 \cdot 10^7$, \textbf{c)} $t = 0.72 \cdot 10^7$ and \textbf{d)} $t = 10^7$. The calculations were performed for the following parameters: $c_0 = 0.003$, $\beta E_{V} = 4.0$, $\beta E_{ES} = 0.0$, $n_{DS} = 5$ and system size $400$ x $400$.}
\label{fig:time}
\end{figure*}

\begin{figure*}[hbt]
 \centering
 \begin{tabular}{|c@{}c|c|}
   \hline
     &  $\beta E_{V} = 3.0$  &  $\beta E_{V} = 4.5$ \\
    \rotatebox[origin=c]{90}{with nucleation}
     &
     \begin{tabular}{@{}c@{}c@{}}
       a)\includegraphics[width=0.22\textwidth]{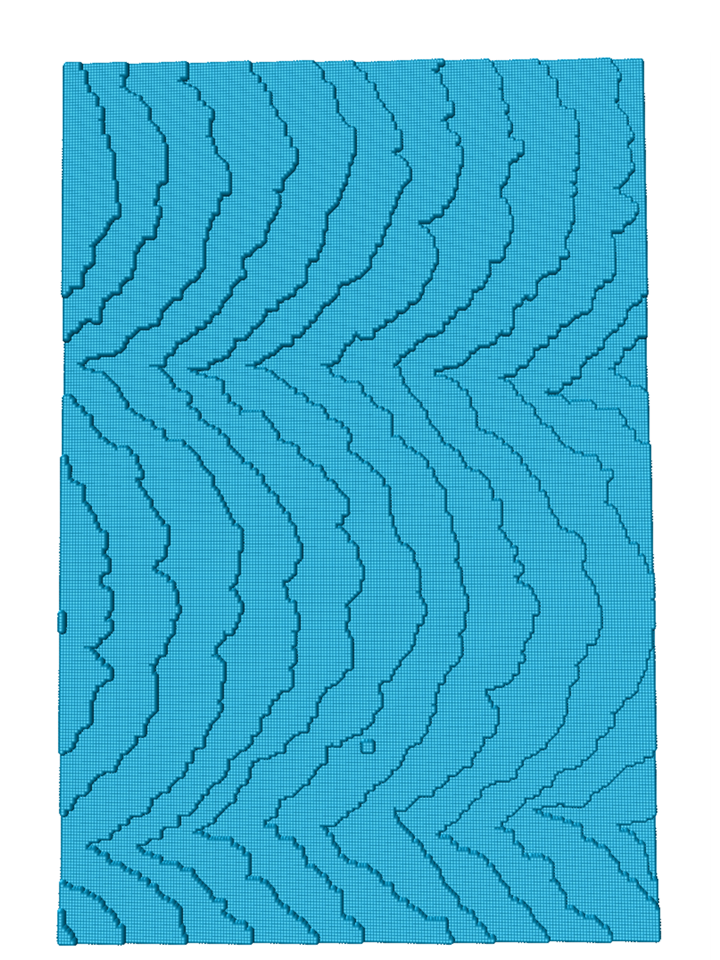} & b)\includegraphics[width=0.22\textwidth]{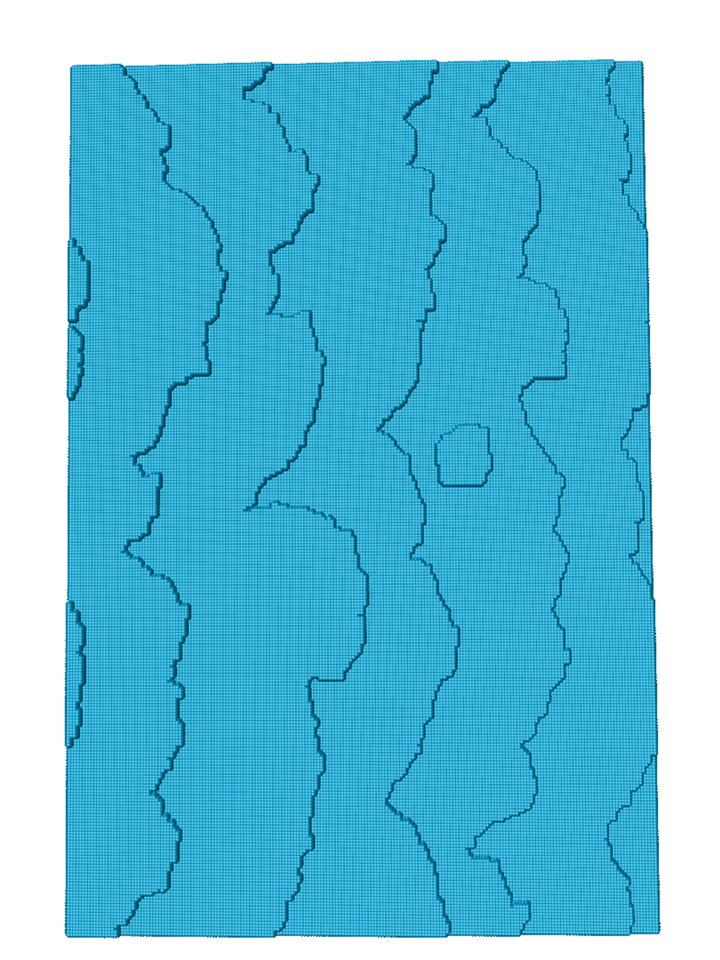}
     \end{tabular}
       &
     \begin{tabular}{@{}c@{}c@{}}
       e)\includegraphics[width=0.22\textwidth]{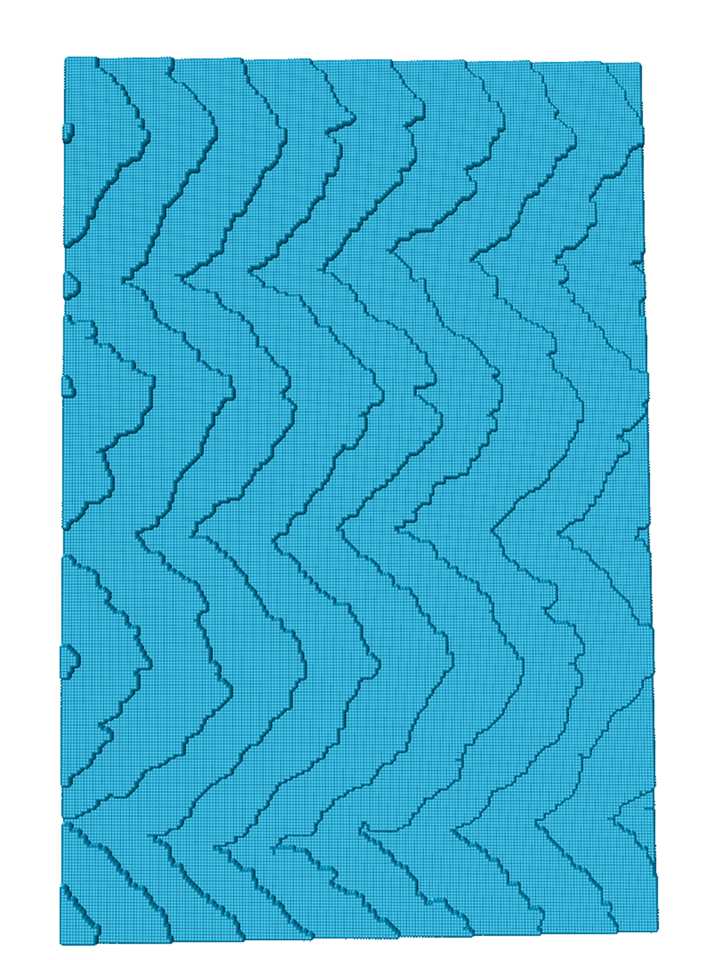} & f)\includegraphics[width=0.22\textwidth]{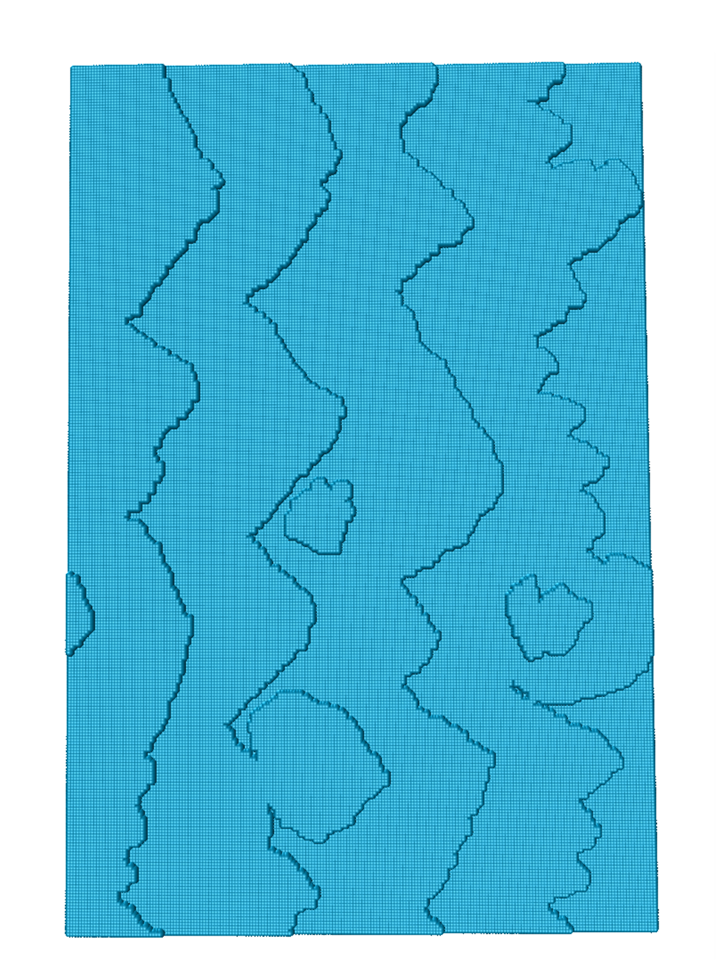}
     \end{tabular} \\
     \hline
     \rotatebox[origin=c]{90}{without nucleation}
     &
      \begin{tabular}{@{}c@{}c@{}}
      c)\includegraphics[width=0.22\textwidth]{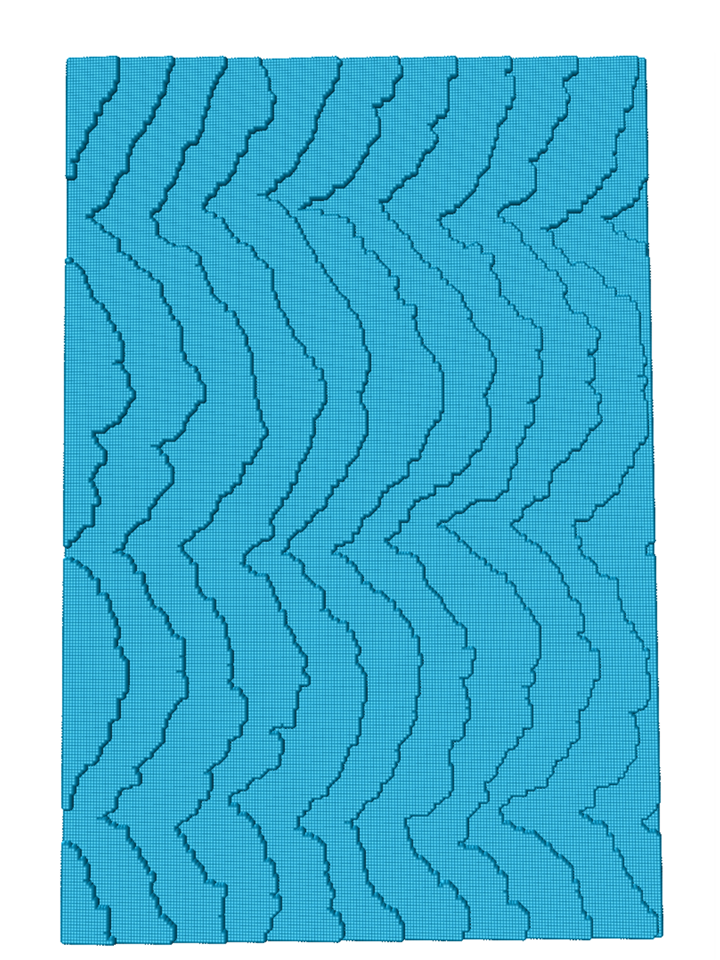} & d)\includegraphics[width=0.22\textwidth]{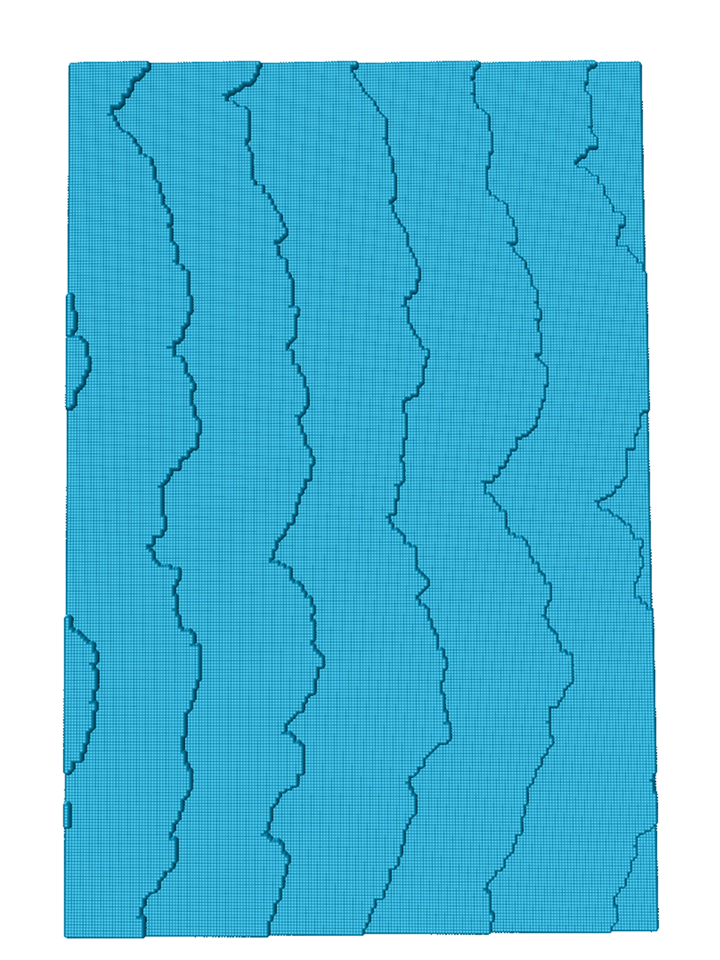} \\
     \end{tabular}
       &
     \begin{tabular}{@{}c@{}c@{}}
    g)\includegraphics[width=0.22\textwidth]{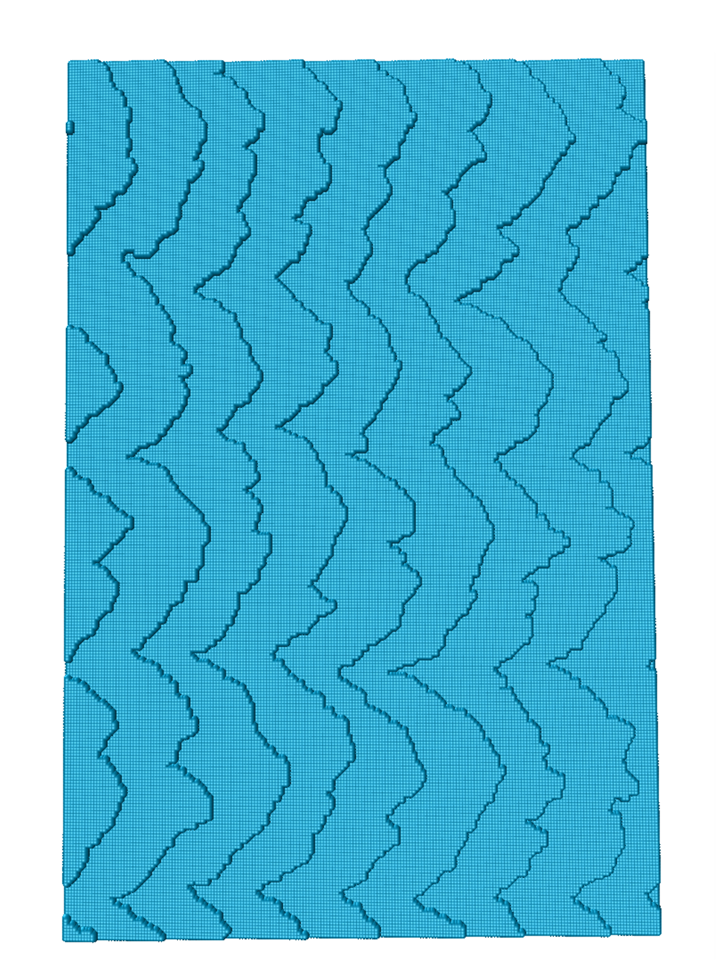} & h)\includegraphics[width=0.22\textwidth]{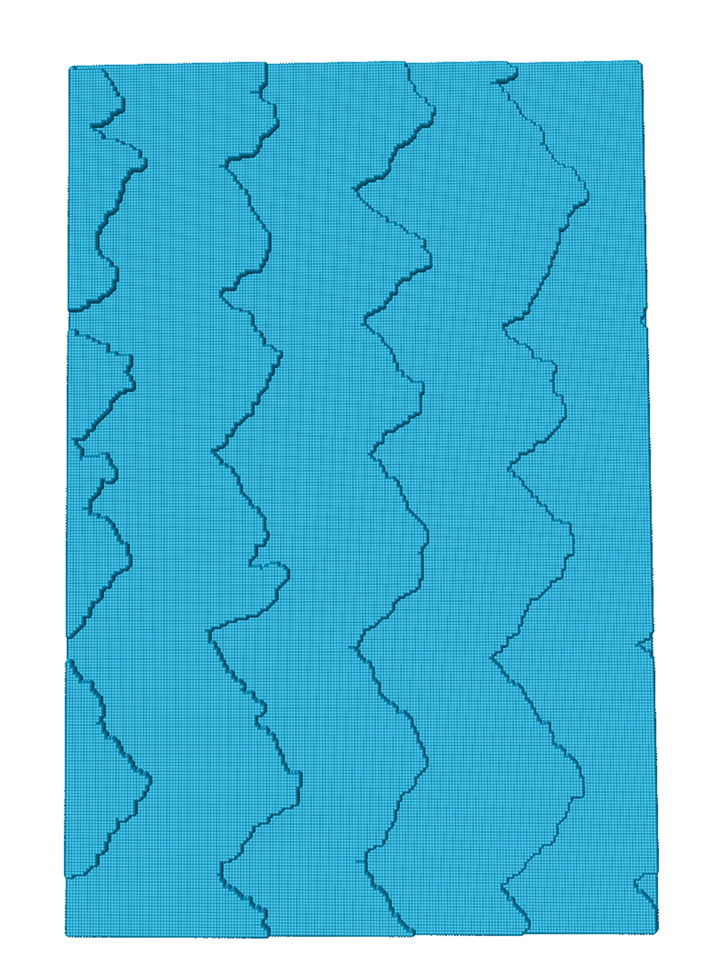} \\
     \end{tabular} \\
  \hline
  \end{tabular} \\
\caption{\textbf{Nucleation of islands.} Comparison of meanders obtained in simulations with allowed (top panel) and forbidden (bottom panel) nucleation for two different potential well depths $\beta E_{V} = 3.0$ (left panel) and $\beta E_{V} = 4.5$ (right panel). In the case of the shallower potential well shown for the two terrace lengths \textbf{(a, c)} $l_0 = 20$ and \textbf{(b, d)} $l_0 = 40$, island formation is obtained when nucleation is allowed, while none is formed when nucleation is forbidden. In the case of $\beta E_{V} = 4.5$, results are shown for \textbf{(e, g)} $l_0 = 25$ and \textbf{(f, h)} $l_0 = 50$ terrace lengths. Island formation and coalescence are only seen for the larger terrace when nucleation is allowed. The calculations were performed for the following parameters: $c_0 = 0.005$, $n_{DS} = 5$, $\beta E_{ES} = 0.0$, $t = 10^6$, and system size $200$ x $300$.}
\label{fig:nucleation}
\end{figure*}

\end{document}